\documentclass[conference]{IEEEtran}
\IEEEoverridecommandlockouts

\usepackage{cite}
\usepackage{amsmath,amssymb,amsfonts}
\usepackage{algorithmicx}
\usepackage{algcompatible}
\usepackage{graphicx}
\usepackage{textcomp}
\usepackage{xcolor}
\usepackage{marvosym}
\usepackage{amsmath}
\usepackage{epstopdf}
\usepackage{subfigure}
\usepackage{algorithm}
\usepackage{makecell}
\usepackage[noend]{algpseudocode}

\def\BibTeX{{\rm B\kern-.05em{\sc i\kern-.025em b}\kern-.08em
    T\kern-.1667em\lower.7ex\hbox{E}\kern-.125emX}}
\begin{document}

\title{Significant-attributed Community Search in Heterogeneous Information Networks\\
}

\author{
\IEEEauthorblockN{Yanghao Liu\textsuperscript{\dag\S}, Fangda Guo\textsuperscript{\dag\Letter}, Bingbing Xu\textsuperscript{\dag}, Peng Bao\textsuperscript{\ddag}, Huawei Shen\textsuperscript{\dag}, Xueqi Cheng\textsuperscript{\dag}}
\IEEEauthorblockA{
\textsuperscript{\dag}\textit{Institute of Computing Technology, Chinese Academy of Sciences, Beijing, China}\\
\textsuperscript{\S}\textit{University of Chinese Academy of Sciences, Beijing, China}\\
\textsuperscript{\ddag}\textit{Beijing Jiaotong University, Beijing, China}\\
\texttt{\small\{liuyanghao19s, guofangda, xubingbing, shenhuawei, cxq\}@ict.ac.cn,} \texttt{\small baopeng@bjtu.edu.cn}}
}

\maketitle

\begin{abstract}
Community search is a personalized community discovery problem aimed at finding densely-connected subgraphs containing the query vertex. In particular, the search for communities with high-importance vertices has recently received a great deal of attention. 
However, existing works mainly focus on conventional homogeneous networks where vertices are of the same type, but are not applicable to heterogeneous information networks (HINs) composed of multi-typed vertices and different semantic relations, such as bibliographic networks. In this paper, we study the problem of high-importance community search in HINs. 
A novel community model is introduced, named heterogeneous significant community (HSC), to unravel the closely connected vertices of the same type with high attribute values through multiple semantic relationships. An HSC not only maximizes the exploration of indirect relationships across entities of the anchor-type but incorporates their significance. To search the HSCs, we first develop online algorithms by exploiting both segmented-based meta-path expansion and significance increment. Specially, a solution space reuse strategy based on structural nesting is designed to boost the efficiency. In addition, we further devise a two-level index to support searching HSCs in optimal time, based on which a space-efficient compact index is proposed. Extensive experiments on real-world large-scale HINs demonstrate that our solutions are effective and efficient for searching HSCs, and the index-based algorithms are 2-4 orders of magnitude faster than online algorithms.

\end{abstract}


\section{Introduction}
\newtheorem{definition}{Definition}
\newtheorem{theorem}{Theorem}
\newtheorem{example}{Example}
\newtheorem{lemma}{Lemma}
\newtheorem{corollary}{Corollary}
\newenvironment{proof}{{\it Proof}{\it :}}{\hfill $\square$\par}
\newenvironment{proof sketch}{{\it Proof Sketch}{\it :}}{\hfill $\square$\par}
Community search is a kind of query-dependent personzlized community discovery problem and aims to find densely-connected subgraphs containing query vertices, has already become an area of high interest\cite{ sun2012mining, cui2014local, huang2015approximate, fang2016effective, fang2017effective,li2018skyline, han2022data}. In order to more accurately express the cohesiveness and specificity, in recent years, studies that incorporate the importance features within the community search has gained a great deal of attention\cite{li2015influential,huang2017attribute,li2017finding,luo2020efficient,zhou2023influential}.
These existing works mainly focus on conventional homogeneous networks where vertices and edges are of the same type, and there is already a relatively mature body of knowledge and solutions. However, due to the differences in underlying structures, models and methods applicable on homogeneous graph are not suitable for heterogeneous graph. To handle more complex network structures, some researchers have shifted their focus to community search in heterogeneous graphs.\cite{hu2019discovering,fang2020effective,qiao2021keyword,jiang2022effective,fang2022cohesive,zhang2021pareto}.

\begin{figure}[t]
    \centering
    \subfigure[An HIN example] { \label{fig:IMDB}
    \includegraphics[width=0.53\columnwidth]{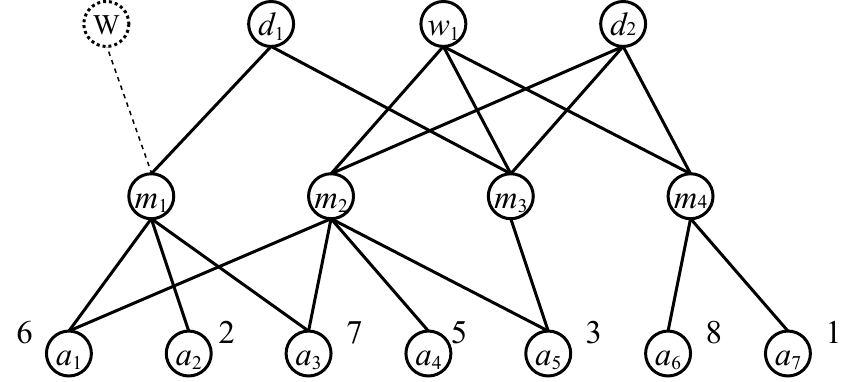}
    }
    \subfigure[Commonly used meta-paths]{ \label{fig:meta-path}
    \includegraphics[width=0.4\columnwidth]{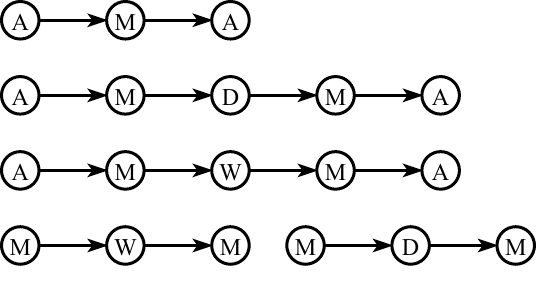}
    }
    \vspace{-10pt}
    \caption{An HIN example and meta-paths (IMDB)}
    \label{fig:IMDB+MP}
    \vspace{-15pt}
\end{figure}

Heterogeneous information networks (HINs) are graph-like structures that represent complex systems consisting of multiple types Lof objects and multiple typed links represent their relationships\cite{sun2011pathsim, shi2016survey}. This kind of network structure can express many real-world scenarios to catch the local dense subgraphs (e.g., social network, academic network, biological network). Incorporating additional attribute information on HIN can enrich the expression of graph relationships further, such as personal information of each user in social networks, citations of papers and the influence of researchers in biological network.
As shown in Fig.\ref{fig:IMDB}, take the classic heterogeneous graph IMDB network\cite{maas2011learning} as an example. It has four vertex types, i.e., author (A), movie (M), director (D), writer (W), the value on each vertex represents its vertex attribute, which means the influence of a movie or person in the IMDB network. And there exist three symmetrical reversible vertex relationships, i.e., an actor is in a movie (A$\rightarrow$M), a director directs a movie (D$\rightarrow$M), a writer writes a movie's script (W$\rightarrow$M) and the inverse relationship of these three relationships, which is called meta-path in HIN area. From the above structure and attribute information, a graph network relationship can be fully expressed.


In this regard, we aim to devise a holistic representation method that takes into account meta-path, graph cohesiveness, and node attribute information. This is to articulate the interrelationships among vertices in a heterogeneous graph, the degree of anchor-type association, and the relative representativeness exhibited by attributes, thus holistically expressing high-importance heterogeneous graph communities. Building upon existing research and our insights, we define this issue as the problem of \textbf{S}ignificant \textbf{A}ttribute \textbf{C}ommunity search over \textbf{H}eterogeneous information network, and we refer to the communities identified by our approach as \textbf{H}eterogeneous \textbf{S}ignificant \textbf{C}ommunity.



\textbf{Applications.} The HSC has extensive application scenarios, listed below are as follows: 
(1) \textit{Social Network Analysis.} In social networks, modeling entities(e.g., users, organizations, events) as vertices, with interactions(e.g., sharing, liking, commenting) as edges and properties(e.g., page reviews, total likes and number of fans) as attributes. Finding HSCs in this context aids in identifying influential social groups and capturing collective opinions.
(2) \textit{Recommendation System.} In e-commerce or content platforms, an HIN is constructed with vertices symbolizing entities like consumers and products, while the edges depict relationships such as purchasing, reviewing, and category affiliation. The vertex attributes encapsulate details such as consumer purchasing power and product sales. Leveraging HINs, with consumers or products as anchor-types, facilitates the extraction of the HSCs, which enable platforms to precisely match products with potential buyers of similar intent and financial capacity, and vice versa, optimizing the precision and efficiency of recommendation systems. 
(3) \textit{Specific Group Discovery.} The HSCs can help identify an influential group within some other specific area. For instance, in widely used scientific collaboration networks, by focusing on specific research domains and influence metrics, one can extract high-quality collaboration circles to pinpoint elite research teams, which can be achieved by setting meta-paths and significance to yield such a dense subgraph. 
In summary, the SACH problem plays a significant role in enhancing network security awareness and governance capabilities, offering considerable commercial and societal benefits.

\textbf{Challenge.} Since the HSC is a kind of community model that comprehensively computes and represents both attributes and structures over HINs, and is attribute-driven, the process of establishing HSC poses the following challenges: 
(1) \textit{Processing the Heterogeneous Structure.} 
Several existential approaches can address the heterogeneity of HINs\cite{sun2011pathsim, dong2017metapath2vec}. However, given the additional attribute constraints in our task scenario and the pursuit of efficient modelling methods, we seek a method that starts with just the query vertex for localized processing and simultaneously advancing the transforming candidate by attributes in the process, which is a extremely hard task.
(2) \textit{Pruning Communities based on Attributes.} Our objective is to create a community emphasizing attribute strengths. In this process, selectively filtering non-dominant attribute vertices without undermining k-cohesiveness presents a significant challenge.
(3) \textit{Optimizing Computational Efficiency.} There exist some straightforward methods; however, due to the complexity of real-world graph data, the construction of HSC may incur significant computational costs, it is a problem to balance accuracy with efficiency during searching HSCs. 
(4) \textit{Establishing Offline Index Structure.} While we can optimize online query algorithms, the inherent computational cost of searching relevant vertices and edges in large scale datasets is inevitable and correlates with the size of the resultant HSCs. In addressing the SACH-Problem, an offline index construction with online querying paradigm is advantageous. However, it remains challenging to construct an efficient and accurate index and facilitate rapid HSC queries based on it.

\textbf{Contributions.} To address the myriad challenges of the HSC problem, we made the following contributions to enhance our research: 
\begin{itemize}
    \item We devise a community search online query algorithm that incorporates vertex significance information, which is effectively identifies and optimizes communities that include significance information. 
    \item Subsequently, we further optimize this algorithm, which includes conducting joint queries using meta-paths and $k$ to narrow down the solution space, as well as employing cache to avoid redundant computations. 
    \item In addition to this, we introduce two full-graph-based indexing and corresponding construction and query algorithms. 
    \item Finally, we conducted comprehensive experiments on our algorithm using four real-world datasets, evaluating both its effectiveness and efficiency.

\end{itemize}

\section{Problem Statement}
In this section, we formally introduce the heterogeneous significant community (HSC) and its search problem.

\subsection{Preliminaries}

\begin{definition}
\text{\textit{\textbf{(HIN).}}} An HIN $G = (V, E, T)$ consists of a vertex set $V$, an edge set $E$, and a vertex type set $T$. Each vertex $v \in V$ belongs to a specific type $t \in T$, and each edge $e \in E$ connects two vertices $u, v \in V$ of different types $(t_u, t_{v}) \in T \times T$ . 
\end{definition}

\vspace{3pt}

\begin{definition}
\label{definition:meta-path}
\text{\textbf{\textit{(Meta-path).}}} Given an HIN $G = (V, E, T)$, a meta-path $\mathcal{P} = (t_1, t_2, \ldots, t_l)$ is a sequence of vertex types that defines the pattern of path instances in $G$, where $\forall i \in \{1, \ldots, l\}$, $t_i \in T$ and $l-1$ is the length of $\mathcal{P}$. 
\end{definition}

\vspace{3pt}

\begin{definition}
\text{\textbf{\textit{(Symmetric Meta-path).}}} A meta-path $\mathcal{P} = (t_1, t_2, \ldots, t_l)$ is symmetric if $\forall i \in \{1, \ldots, l\}$, $t_i=t_{l-i+1}$.
\end{definition}
\vspace{3pt}

Since we focus on finding communities of vertices with the anchor-type, all the meta-paths mentioned later are symmetric.

\vspace{3pt}
\begin{definition}
\label{defin:Connectivity}
\text{\textbf{\textit{($\mathcal{P}$-Connected vertices}}).} For vertices $u, v \in V$, one given meta-path $\mathcal{P}$, if both $u$ and $v$ belongs to the instance of $\mathcal{P}$, $u$ is considered a $\mathcal{P}$-neighbor of $v$. If there exists a chain of vertices between $u$ and $v$ such that each vertex in the chain is a $\mathcal{P}$-neighbor of its adjacent vertices, then $u$ and $v$ are considered $\mathcal{P}$-connected.
\end{definition}
\vspace{3pt}
\begin{example}
For the IMDB network depicted in Fig.~\ref{fig:IMDB}, consider $\mathcal{P}_1$=\textit{(AMA)} presented in Fig.\ref{fig:meta-path} as the specified symmetrical meta-path. 
A meta-path instance 
$\textit{p}=a_1 \rightarrow m_1 \rightarrow a_2$ subsists, 
denoting vertices $a_1$, $a_2$ interconnected via $\mathcal{P}_1$. 
Correspondingly, $a_2$ and $a_5$ are also $\mathcal{P}_1$-connected. While no direct meta-path instances exist between them, i.e., $a_2$ is not a $\mathcal{P}_1$-neighbor to $a_5$, a chain of vertices $a_2 \rightarrow m_1 \rightarrow a_1 \rightarrow m_2 \rightarrow a_5$ does exist. 
In other words, the chain is concatenated from two meta-path instances.
\end{example}
\vspace{3pt}


\subsection{Problem Definition}

\begin{definition}
\label{anchor-type k-core}
\text{\textbf{\textit{(Anchor-type $k$-core).}}} Given an HIN $G=(V,E,T,\phi,\psi)$ where $T=\{t_1, t_2, \ldots, t_n\}$, one may elect $t_{anchor} \in T$ to serve as the anchor-type, functioning as the principal research type. Given a meta-path $\mathcal{P}$, if $\exists G' \subseteq G$, $\forall v \in G'$ and $t_v=t_{anchor}$, $\alpha(v, G'_\mathcal{P}) \geq k$, here $\alpha(v, G'_\mathcal{P})$ indicates the number of $\mathcal{P}$-neighbors of $v$ in $G'_\mathcal{P}$, $G'$ is deemed to satisfy the anchor-type $k$-core, denoted as $Core(G', t_{anchor}, \mathcal{P})$. 
\end{definition}

\vspace{3pt}
\begin{definition}
\text{\textbf{\textit{($k, \mathcal{P}$)-core\cite{fang2020effective}.}}} Given an HIN $G$, a positive integer $k$, and a meta-path $\mathcal{P}$. Solely focusing on the anchor-type vertices, the Basic($k$, $\mathcal{P}$)-core represents the maximal $\mathcal{P}$-connected vertex set denoted as $\mathcal{B}_{k, \mathcal{P}}$ where each $v$ $\in$ $\mathcal{B}_{k, \mathcal{P}}$ meets the condition $\alpha(v, \mathcal{B}_{k, \mathcal{P}})$ $\geq$ $k$.
\end{definition}

\vspace{3pt}

\begin{example}
\label{example:without attribute CS}
Given a positive integer $k$=4, let $\mathcal{P}_1$=(\textit{AMA}), $\mathcal{P}_2$=(\textit{AMDMA)}, $\mathcal{P}_3$=(\textit{AMWMA}). According to the HIN $G$ displayed in Fig.~\ref{fig:IMDB+MP} and the aforementioned query conditions, no $\mathcal{B}_{4, \mathcal{P}_1}$ will materialize as the vertex set constituted by instances of $\mathcal{P}_1$ cannot construct a 3-core. Furthermore, $\mathcal{B}_{4, \mathcal{P}_2}$=$\{a_1, a_3, a_4, a_5, a_6, a_7\}$ , as these points can construct a 5-clique based on the meta-path instances formed by $\mathcal{P}_2$, satisfying the 5-core definition while inherently being a 4-core. Identically, $\mathcal{B}_{4, \mathcal{P}_3}$=$\{a_1, a_3, a_4, a_5, a_6, a_7\}$. In fact, the maximum value of  cohesiveness constraints is 5 when provided with meta-path $\mathcal{P}_2$ or $\mathcal{P}_3$, and they all yield identical results. i.e., $\mathcal{B}_{4, \mathcal{P}_2}$=$\mathcal{B}_{4, \mathcal{P}_3}$=$\mathcal{B}_{5, \mathcal{P}_2}$=$\mathcal{B}_{5, \mathcal{P}_3}$. This demonstrates that merely considering $k$-core and meta-path constraints is insufficient to derive a more accurate and representative community result. Conjoining this with the network structure of the real world, more graph information deserves consideration.
\end{example}
\vspace{3pt}

In fact, each type of vertices has its own significance. For example, in the IMDB network, the significance is often used to indicate the influence of actors (A-type vertices) or the ratings of movies (M-type vertices), etc. Assume we aim to identify groups of highly popular actors who have co-starred in movies; the community model needs to take into account and integrate several constraints, specifically, acting experiences, collaborative relationships, actor popularity.

\vspace{3pt}
\begin{definition}
\label{definition:Attribute of Community}
\text{\textbf{\textit{(Significance of Community).}}} Given a community \(\mathcal{H}\), let $Sig(v)$ symbolizes the significance of vertex $v$, then the significance of community embodies the minimum significance of vertices within the community, i.e., \(f(\mathcal{H})\)\(=\)\(min\{Sig(v) | v\in \mathcal{H}\}\). 
\end{definition}


\vspace{3pt}
\textbf{SACH-Problem.} Given an HIN \(G=(V,E,T,\varphi,\phi)\), a non-negative integer \(k\), a symmetric meta-path \(\mathcal{P}\), and a query vertex \(q \in V\), we aim to find a vertex set \(\mathcal{C}\) in graph \(G\) with vertex type \(\varphi(q)\) that maximizes the community significance \(f(\mathcal{C})\) and satisfies the following conditions:
\begin{itemize}
\item Connectivity: \(\mathcal{C}\) contains \(q\) and is $\mathcal{P}$-connected to any vertex \(v \in \mathcal{C}\).
\item Cohesiveness: $\forall$\(v \in \mathcal{C}\), \(\alpha(v,\mathcal{C}) \geq k\).
\item Maximality: There does not exist another vertex set \(\mathcal{C}'\) that satisfies the above two conditions, \(\mathcal{C} \subseteq \mathcal{C}'\), and \(f(\mathcal{C}') = f(\mathcal{C})\).
\end{itemize}

We refer to a community that satisfies the above conditions as a Heterogeneous Significant Community (HSC).
Correspondingly, we define the method for searching HSC problems as the \textbf{S}ignificant-\textbf{A}ttributed \textbf{C}ommunity \textbf{S}earch over \textbf{H}eterogeneous network Problem (SACH-Problem). The SACH-Problem aims to find tight communities of the same type as the query vertex \(q\), cohesive with respect to meta-paths, and with numerical significance that carry the most weight. 
The symmetric meta-path is applied to ascertain the vertex type and the query vertex in the target community. Following this, the connectivity condition ensures that other vertices are associated with \(q\) based on the meta-path restriction. 
The cohesiveness condition guarantees the closeness of the community. A larger \(k\) suggests more meta-paths starting from \(q\) with different end vertices. It could also imply that the vertices in the community have a higher correlation with \(q\) based on given meta-paths.
Finally, the maximality condition guarantees the community, as depicted by the vertex significance value, holds the most influence; i.e., no other community larger than the result possesses the same or higher community value. Naturally, the resulting community is characterized as an HSC. 
Additionally, we propose a lemma regarding the nested relationship between significance and HSC.

\vskip 3pt
\begin{lemma}
\label{lemma:attribute value relationship}
Given an HIN G, a vertex $u$ and two HSCs $\mathcal{C}_1$ and $\mathcal{C}_2$, where $\mathcal{C}_2 \subseteq \mathcal{C}_1$ and $\mathcal{C}_1 = \mathcal{C}_2 \cup \{u\}$, if the significance value of vertex $u$ is less than the community value of $\mathcal{C}_2$, i.e., $Sig(u) < f(\mathcal{C}_2)$, then $f(\mathcal{C}_1) < f(\mathcal{C}_2)$.
\end{lemma}
\vskip 3pt

\begin{proof}
According to Definition~\ref{definition:Attribute of Community}, the significance satisfy the following formula: 
\begin{equation}
f(\mathcal{C}_1)=\min\{Sig(v) \mid v \in \mathcal{C}_1\} \leq Sig(u) < f(\mathcal{C}_2). 
\end{equation}
Therefore, Lemma~\ref{lemma:attribute value relationship} is proved.
\end{proof}
\vskip 3pt

\begin{example}
When incorporating the significance of anchor-type vertices, let $k$=4 and meta-path \(\mathcal{P}\)=(\textit{AMDMA}) in Fig.~\ref{fig:IMDB}. The result community in Example~\ref{example:without attribute CS}, denoted as $\mathcal{C}_1$, is equal to $\mathcal{B}_{4,\mathcal{P}}=\{a_1, a_3, a_4, a_5, a_6, a_7\}$, whose significance is $f(\mathcal{C}_1)=1$. However, we can identify another community, $\mathcal{C}_2=\{a_1,a_3,a_4,a_5,a_6\}$, which is a subset of $\mathcal{C}_1$ of significance $f(\mathcal{C}_2)=3$. Thus, a more precise and high-quality community is derived.

\end{example}

\section{Online Algorithms}


\subsection{Basic Online Algorithm}\label{AA}
Given an HIN $G=(V,E,T,\phi,\psi)$, a positive integer \(k\), a meta-path \(\mathcal{P}\), and a query vertex \(q\), the algorithm aims to find a target community, denoted as vertex set \(\mathcal{C}\), that satisfies the criteria of the HSC. A straightforward approach is as follows: First, construct a simple graph \(G_{\mathcal{P}}\) consisting of vertices with type \(\phi(q)\) based on the meta-path \(\mathcal{P}\). Then, compute the maximum connected \(k\)-core of \(G_{\mathcal{P}}\) that contains \(q\), denoted as \(G_{\mathcal{P},k}\). Next, iterate over \(G_{\mathcal{P},k}\) and, in the \(t\)-th iteration, remove the vertex with the smallest significance value from \(G_{\mathcal{P},k}^{t-1}\) and delete any vertices that no longer satisfy the \(k\)-core constraint. If, after the removal operation, \(G_{\mathcal{P},k}^t\) contains no vertices, output all the vertices in \(G_{\mathcal{P},k}^{t-1}\) as the target community \(\mathcal{C}\). The specific process is relatively straightforward. Due to space constraints, the specific algorithm will not be given here. It can be referred to the full version of this paper for detailed content.

The fundamental premise of the algorithm's core philosophy commences with the conversion of a heterogeneous graph into a homogeneous one via the meta-path $\mathcal{P}$. It iteratively peels vertices with the least significance values, concurrently sustaining $k$-core and encompassing the query vertices $q$. Ultimately, it yields the largest-scale community that complies with both structural and significance stipulations. At face value, this approach may appear insufficiently efficient, primarily due to the following three considerations:
\begin{itemize}
\item In a large-scale HIN, the quantity of meta-path instances and the length of the meta-path amplify significantly. The former may exhibit an exponential growth trajectory influenced by the latter, leading to a relatively lower time and space efficiency of rudimentary processing.
\item  Not all vertices, homologous to the query vertex $q$, exhibit a $\mathcal{P}$-connected relationship with $q$ within the HIN. In the straightforward manner, these vertices are all accounted for during the construction of the homogeneous graph and processing of the meta-path instance. This inclusion could considerably escalate the time and space demands of the algorithm.
\item The processing of meta-path instances can encompass repetitions or redundancies. For instance, two different meta-paths are formed by exchanging the head and tail vertices of a meta-path instance, the purpose of enumerating them, however, is the same. Both means to establish an edge between the two vertices when inducing the homogeneous graph.
\end{itemize}

Given the limitations of the top-down vertex deletion approach, we pivot to a bottom-up methodology. Specifically, the concept of expansion from the query node and other pertinent target vertices. Intuitively, given a query vertex, only need to consider the vertex itself and its associated target vertex expansion in accordance with the meta-path instance to acquire a community. During this process, the constraint of significance is also contemplated to procure the target community. The optimization algorithm named \textbf{QHSC} (Query Heterogeneous graph Significant Community) is proposed based on the notion of segmental extension of meta-path instances and the principle of limiting the extension of originating vertices.

\subsection{Optimization Algorithm QHSC}
In practical terms, each meta-path represents a node relationship. Therefore, our optimization algorithm dissects the meta-path into segments, beginning expansion from the query vertex $q$ in batches for each segment of the meta-path, denoted as $\mathcal{P}$=$\{\mathcal{s}_1, \mathcal{s}_2, \ldots, \mathcal{s}_l\}$, where $l$ indicates the length of $\mathcal{P}$. Notably, our objects of study in the meta-path are symmetric structures, prompting us to partition the meta-path into two symmetric sections. 
On the other hand, we can naturally regard the heterogeneous graph in the form as a tree. The edge relationships between different types of vertices can be correspondingly constructed as parent-child relationships or ancestor-descendant relationships. 
Based on the above observations,  in conjunction with the concept of meta-path splitting, we amalgamate the procedures of constructing the homogeneous graph from both bottom to top and top to bottom. Specifically, we initially identify all the highest ancestor nodes of the current anchor-type vertex in the current meta-path, subsequently discovering all the anchor-type vertices from these ancestor nodes. For these vertices, it suffices to establish edges among them, excluding the current node itself, facilitating the rapid acquisition of the homogeneous graph. Based on these insights, we present the process of the \textit{GetHomoGraph} function.

\begin{algorithm}
\small
\caption{GetHomoGraph}
\label{algo:getHomoGraph}
\begin{algorithmic}[1]
\State \textbf{Function} \textit{GetHomoGraph($G, \mathcal{P}, v$)}
\State \textbf{Input}: The HIN $G=(V,E,T,\phi,\psi)$, meta-path $\mathcal{P}$ consists of $\mathcal{s}$=$\{\mathcal{s}_1, \mathcal{s}_2, \ldots, \mathcal{s}_l\}$, starting vertex $q$ and the anchor-type of $q$ denoted as $t_{anchor}$
\State \textbf{Output}: The result homogeneous graph $\mathcal{H}$  relative to
\State $\mathcal{H} \gets \varnothing$, $Q \gets \varnothing$, $Q.add(v)$,$Exclude \gets \varnothing$
\State $mid \gets \frac{l}{2}$, $t_{lead} \gets$ lead vertex type of $\mathcal{P}$
\State $\mathcal{s}_{upon} \gets \{\mathcal{s}_1, \ldots, \mathcal{s}_{mid}\}$
, $\mathcal{s}_{down} \gets \{\mathcal{s}_{mid+1}, \ldots, \mathcal{s}_l\}$
\While{$Q \neq \varnothing$}
    \State $Vertices \gets \varnothing$, $Anchor \gets \varnothing$
    \State $cur \gets Q.poll()$, $Exclude.$add($cur$)
    \State $Vertices \gets Expand(cur, \mathcal{s}_1, Vertices)$
    \State $Anchor \gets$ all subordinate vertices in type $t_i$ of $Vertices$
    \State $Anchor \gets Anchor \setminus Exclude$
    \For{each $anchor \in Anchor$}
        \State link $cur$ and $anchor$ in $\mathcal{H}$
        \State $Q.$push($anchor$)
    \EndFor
\EndWhile
\State\Return $\mathcal{H}$
\Statex
\State\textbf{Function} \textit{Expand($\mathcal{s}_i, vertex, Vertices$)}
\State $Nei \gets vertex.\mathcal{s}_i$-$neighbors$
\For{each $u \in Nei$}
    \If{$u.type=t_{lead}$}
        \State $Verticecs.$add(u)
    \EndIf
    \If{$\mathcal{s}_{i+1} \in \mathcal{s}_upon$}
        \State $Vertices.$addAll($Expand(\mathcal{s}_{i+1}, u, Vertices$))
    \EndIf
\EndFor
\State \Return $Vertices$
\end{algorithmic}
\end{algorithm}

Algorithm~\ref{algo:getHomoGraph} outlines the procedure for generating the homogeneous graph linked to $q$, predicated on the provided meta-path $\mathcal{P}$. We segment $\mathcal{P}$ into a set $\mathcal{s}$=$\{\mathcal{s}_1, \mathcal{s}_2, \ldots, \mathcal{s}_l\}$ as the input.  The algorithm starts with some initialization procedures and we divide $\mathcal{s}$ into two halves, namely $\mathcal{s}_{upon}$ and $\mathcal{s}_{down}$ as we have stipulate that $\mathcal{P}$ is symmetric (lines 4-6).  Within each iteration, we pop vertex $cur$ from $Q$ and add it to $Exclude$  (lines 8-9). We then recursively invoke the function $Expand(\mathcal{s}_i,vertex,Vertices)$ to obtain the highest ancestor nodes and store them into $Vertices$. Within the Expand function, we initially secure all neighboring nodes of the input $v$ based on $\mathcal{s}_i$ and assess the nodes' roles. If a neighbor coincides with the highest ancestor node type within the algorithm's meta-path input, these nodes are retained. Conversely, if they do not fulfill the conditions of the highest ancestor node type, the recursive search persists (lines 22-29). After the recursion yields results, we incorporate all anchor-type subordinate vertices present within $Vertices$ into $Anchor$. Then we exclude vertices that also in $Exclude$ from $Anchor$, subsequently adding an undirected edge $(cur \rightarrow anchor)$ to $\mathcal{H}$ for each remaining item $anchor$, incidentally, add $anchor$ into queue $Q$ and concluding the current iteration (lines 12-16). Finally we get the result homogeneous graph $\mathcal{H}$.  

\vspace{3pt}
\begin{lemma}
\label{cor:GetHomoGraph}
The vertex set obtained by \textit{GetHomoGraph} is a superset of the final HSC's vertex set.
\end{lemma}
\vspace{3pt}

\begin{proof}
Denote the final HSC as $\mathcal{C}$ and the result obtained by \textit{GetHomoGraph} as $S$. 
Since we first obtain all the highest ancestor vertices of $q$, and then get all the anchor-type vertices from these ancestor vertices, according to \textit{Definition} \ref{defin:Connectivity}, all the vertices in $S$ are $\mathcal{P}$-connected with $q$.
Suppose $\exists v \in \mathcal{C} \cap v \notin S$, given that $v \in \mathcal{C}$, $v$ must be $\mathcal{P}$-connected with $q$, which implies that $v$ must be in $S$, leading to a contradiction. Subsequently, $\mathcal{C}$ is obtained nonincreasingly on $S$, therefore the lemma is proven.
\end{proof}
\vspace{3pt}

\begin{figure}[t]
    \centering
    \includegraphics[width=0.9\columnwidth]{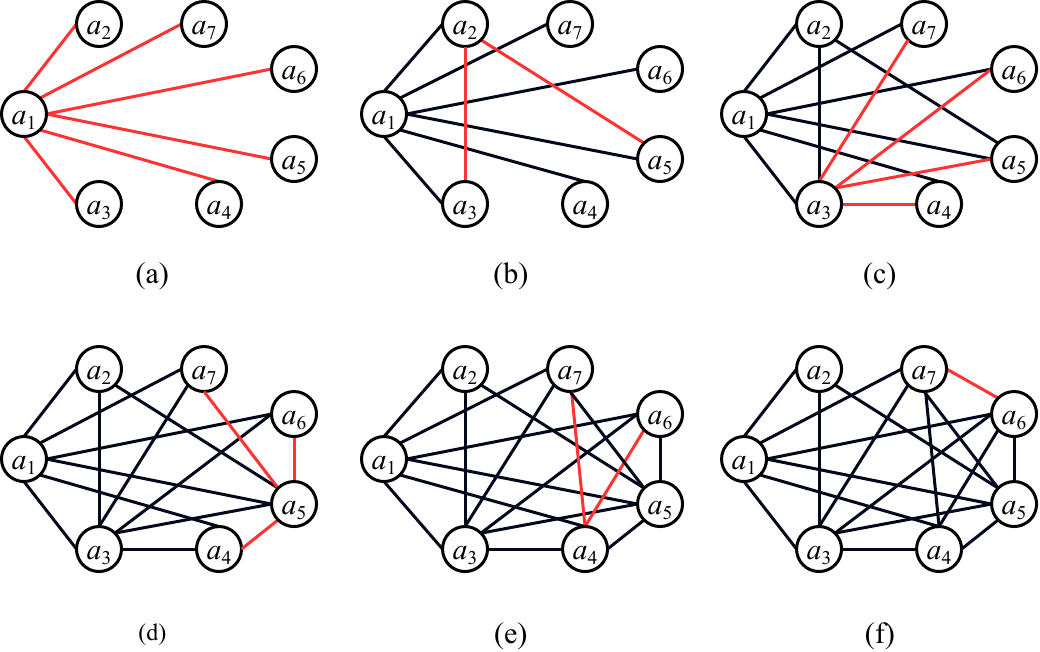}
    \vspace{-3pt}
    \caption{Process of obtaining the homogeneous graph}
    \vspace{-15pt}
    \label{fig:getHomoGraph}
\end{figure}

\begin{example}
Fig.~\ref{fig:getHomoGraph} demonstrates an execution of Algorithm~\ref{algo:getHomoGraph}. Taking Fig.~\ref{fig:IMDB} as input $G$, assume $\mathcal{P}=\text{(\textit{AMDMA})}$, $\mathcal{s}=\{\mathcal{s}_1,\mathcal{s}_2,\mathcal{s}_3,\mathcal{s}_4\}$ and $q$=$a_1$. Initially, we invoke the function $Expand(\mathcal{s}_1, a_1, Vertices)$, which results in $Vertices$=$\{d_1,d_2\}$. Then we find that all subordinate vertices thereof constitute $Anchor$=$\{a_1,a_2,a_3,a_4,a_5,a_6,a_7\}$, following the exclusion process and edge addition in $\mathcal{H}$, we obtain $\mathcal{H}$ as shown in Fig.~\ref{fig:getHomoGraph}(a), and sequentially incorporate vertices $a_2, a_3, ..., a_7$ into $Q$. Subsequently, $Q$ pops up $a_2$ for subsequent operations, which are shown in Fig.~\ref{fig:getHomoGraph}(b) to Fig.~\ref{fig:getHomoGraph}(f), and will not be elaborated here.
\end{example}

Upon devising the method for generating a homogeneous graph, the subsequent step entails deriving the resultant community through structural and significance-based pruning.
In light of our analysis of the requirements for constructing significant communities in heterogeneous graphs, the process of constructing HSC $\mathcal{C}$ can be conceptualized as an iterative process that alternates between checking for optimality in terms of $f(\mathcal{C})$ and $k$-cohesiveness. Inspired by the round-robin approach, the algorithm prioritizes processing nodes with more representative significance to examine their cohesiveness properties. Once a set of candidate communities is acquired, a unified post-processing step is performed to maximize the quality of the resulting communities. 
Based on these analysis, we propose the Query Heterogeneous Significant Community (QHSC) algorithm.

\begin{algorithm}
\small
\caption{QHSC}
\label{algo:QHSC}
\begin{algorithmic}[1]
\State \textbf{Input:} HIN $G = (V, E, T, \varphi, \psi)$, positive integer $k$, meta-path $\mathcal{P}$, query vertex $q$
\State \textbf{Output:} Vertex set $\mathcal{C}$ consisting of HSC requirements
\State $G_P \leftarrow GetHomoGraph(G, P, q)$
\State $G_{P, k} \leftarrow ComputeKCore(G_P, k, q)$
\State $S_{order} \leftarrow$ sort $Sig(v)$ for all $v \in G_{P, k}$ in ascending order 
\While{$S_{order} \neq \varnothing$}
    \State $Delete \leftarrow \varnothing$
    \State $Temp \gets$ minimum vertices of $G_{P, k}$
    \While{$Temp \neq \varnothing$}
        \State $cur \leftarrow Temp.pop()$
        \If{$cur$ is $q$}
            \State $H \leftarrow ComputeKCore(G_P, k, q)$
            \State $\mathcal{C} \leftarrow H \cup Delete$
            \State \Return $\mathcal{C}$
        \EndIf
        \For{each $u$ in $Neighbor(cur, G_{P, k})$}
            \State $\alpha(u, G_{P, k}) \leftarrow \alpha(u, G_{P, k}) - 1$
            \If{$\alpha(u, G_{P, k}) < k$}
                \State $Temp.add(cur)$
            \EndIf
        \EndFor
        \State $S_{order}.remove(cur)$
        \State remove $cur$ and its associated edges from $G_{P, k}$
        \State $Delete.add(cur)$
    \EndWhile
\EndWhile
\State \Return $\mathcal{C}$
\end{algorithmic}
\end{algorithm}

Algorithm~\ref{algo:QHSC} shows the process of QHSC, which accepts $G$, $k$, $\mathcal{P}$, $q$ as the input and returns an HSC $\mathcal{C}$(lines 1-2). 
Initial, retrieves the homogeneous graph $G_\mathcal{P}$ by calling the function \textit{GetHomoGraph} and computing its corresponding core $G_{\mathcal{P}, k}$ (lines 3-4). Subsequently, we sort all vertices in $G_{\mathcal{P}, k}$ in ascending order based on their significance, which supports us to alternate between significance and structural queries (lines 5-6). Within each iteration, we process vertices with minimum significance (lines 7-8). Afterwards, get the top vertex $cur$ and judge if $cur=q$, we can directly get the result HSC (lines 11-14). Otherwise, continue to alternately judge significance and structure, iteratively deleting the vertex with minimum significance until it no longer satisfies the $k$-core constraint (lines 15-21). Finally, we get the result HSC (line 22).

\vspace{3pt}
\begin{lemma}
    Algorithm~\ref{algo:QHSC} can yield the exact solution of $q$.
\end{lemma}
\vspace{3pt}

\begin{proof}
According to \textit{Lemma}~\ref{cor:GetHomoGraph}, the set got from \textit{GetHomoGraph} is the entire potential vertex set of the HSC. In Algorithm~\ref{algo:QHSC}, we ensure that the vertex with the minimum significance, whose iterative deletion does not affect the $k$-core, is removed. Therefore, the final result must satisfy the HSC requirement.
\end{proof}
\vspace{3pt}

\subsection{Advanced Online Algorithm AQHSC}
The Advanced QHSC (AQHSC) algorithm is formulated by exploiting the characteristics of the homogeneous graph, induced by the meta-path, coupled with the cohesiveness inherent to the $k$-core based on the QHSC algorithm, we begin with the foundational observations.

\vskip 3pt
\begin{lemma}
\label{meta-path nesting}
Given symmetric meta-path $\mathcal{P}$ and its sub meta-path $\mathcal{P}'$, i.e., $\mathcal{P}' \subseteq \mathcal{P}$. The anchor-type vertices that are connected by $\mathcal{P}'$ are inevitably connected by $\mathcal{P}$.
\end{lemma}
\vskip 3pt
\begin{proof}
Given an HIN $G$, a symmetric meta-path $\mathcal{P}$ and its sub-meta-path $\mathcal{P}' \subseteq \mathcal{P}$, and two vertices $v_1, v_2\in G$. Per \textit{Definition}~\ref{definition:meta-path}, if $v_2$ is reachable from $v_1$ through $\mathcal{P}'$, then $v_2$ is also reachable from $v_1$ via a segment of $\mathcal{P}$.
The addition of supplementary meta-path segments, not present in $\mathcal{P}'$, can be done by iteratively traversing the repeated paths in the worst case, given that the meta-path instance is exhaustive in our context, there exists at least one strategy can invariably satisfy these conditions.
Consequently, the accessibility of the reverse direction is equivalent on both $\mathcal{P}$ and $\mathcal{P}'$, which is guaranteed by the symmetry of the specified meta-paths in our study. 
\end{proof}
\vskip 3pt

\begin{lemma}
\label{graph nesting}
With the same give conditions of \textit{Lemma}~\ref{meta-path nesting}.
Let $G_\mathcal{P}$ and $G_\mathcal{P}'$ symbolize the homogeneous graphs containing $q$ induced by $\mathcal{P}$ and $\mathcal{P}'$, respectively. Consequently, $G_\mathcal{P}'$ is a subgraph of $G_\mathcal{P}$, represented as $G_\mathcal{P}' \subseteq G_\mathcal{P}$.
\end{lemma}
 \vskip 3pt
\begin{proof}
Given that \( \mathcal{P}' \subseteq \mathcal{P} \) and invoking \textit{Lemma}~\ref{meta-path nesting}, all \( \mathcal{P}' \)-neighbors of \( q \) must also be \( \mathcal{P} \)-neighbors. This implies that the vertices present in the homogeneous graph \( G_\mathcal{P}' \) are also contained within \( G_\mathcal{P} \), denoted as \( G_\mathcal{P}' \subseteq G_\mathcal{P} \), Q.E.D.
\end{proof}
\vskip 3pt
\begin{lemma}
\label{k-core nesting}
For any HIN $G$ adhering to anchor-type $k$-core under the meta-path $\mathcal{P} $, it is inherent that $\forall i \in \left[ 1, k\right]$, $G$ satisfies anchor-type $i$-core with consistent $\mathcal{P}$. i.e., the presence of a $(k, \mathcal{P})$-core implies the existence of a $(i,\mathcal{P})$-core within $G$.
\end{lemma}

\vskip 3pt

\begin{proof}
According to \textit{Definition}~\ref{anchor-type k-core}, the set $\mathcal{S}$ consisting of anchor-type vertices that constitute a $k$-core via $\mathcal{P}$, i.e., $\forall v \in \mathcal{S}$, $\forall i \in \left( 0, k\right]$, $\alpha(v,G) \geq k \geq i$. Q.E.D.
\end{proof}
\vskip 3pt

According to the conclusion in \textit{Lemma}~\ref{graph nesting} and \textit{Lemma}~\ref{k-core nesting}, it is plausible to curtail the size of the solution space for sub-meta-path queries by confining the exploration to the solution space of the parent meta-path. Likewise, the $k$-cohesiveness query can potentially optimize the solution space from smaller $k$, thereby enabling complex joint queries. These observations lay the groundwork for the introduction of our advanced algorithm.

\begin{algorithm}[t]
\small
\caption{AQHSC}
\label{algo:Advanced}
\begin{algorithmic}[1]
\State \textbf{Input:} HIN $G = (V, E, T, \varphi, \psi)$, positive integer $k$, meta-path $\mathcal{P}$, sub-meta-path $\mathcal{P}'$, query vertex $q$.
\State \textbf{Output:} Vertex set $\mathcal{C}$ consisting of HSC.
\State $\mathbb{R} \gets QHSC(G, k, q, \mathcal{P})$
\State $S \gets $all vertices that are $\mathcal{P}$-connected with $q$
\State $S=S \cap \mathbb{R}$
\State Initialize $Visited\gets\varnothing$, $Adj \gets \varnothing$  
\State $Adj \gets Adj.put(q, S)$
\State $Adj \gets$ \textit{ComputeHomoWithCache}$(S, Visited, \mathbb{R}, Adj)$
\State $\mathcal{C} \gets$ tail treatments result of $Adj$ 
\State \Return $\mathcal{C}$
\Statex
\State \textbf{Function} \textit{ComputeHomoWithCache($S, Visited, \mathbb{R}, Adj$)}
\State Initialize $size\gets 0$, $\mathcal{P}NbSet\gets\varnothing$
\For{each $v$ $\in$ $S$}
    \State $size\gets size+1$
    \If{$Visited.contains(v)$}
        \If{$size \neq S.length$}
            \State continue
        \EndIf
        \State \textit{ComputeHomoWithCache}($\mathcal{P}NbSet, Visited, \mathbb{R}, Adj$)
    \EndIf

    \State $Nei \gets \mathcal{P}$-neighbor vertices of $v$
    \State $Nei \gets Nei \cap \mathbb{R}$
    \State $Visited \gets Visited.add(v)$
    \State $\mathcal{P}NbSet \gets \mathcal{P}NbSet.add(Nei)$
    \State $Adj \gets Adj.put(v, Nei)$
    \If{$size=S.length$}
        \State \textit{ComputeHomoWithCache}($\mathcal{P}NbSet, Visited, \mathbb{R}, Adj$)
    \EndIf
\EndFor
\State \Return $Adj$
\end{algorithmic}
\end{algorithm}

As delineated in Algorithm~\ref{algo:Advanced}, we leverage the pre-existing solution space to simplify computations. Initially, we identify  the $\mathcal{P}$-neighbors of $q$ and intersect them with the pref-existing solution space $\mathbb{R}$. 
According to \textit{Lemma}~\ref{graph nesting}, vertices absent from the solution space of parent meta-path are invariably absent from those of the sub-meta-path, allowing us to eliminate certain vertices.
In proceeding, we use $Visited$ to avoid redundant operations and the map $Adj$ to store the community structure (line 6), which stores the key-value pair $(q, S)$ in $Adj$ (line 7). Afterwards, we call the function \textit{ComputeHomoWithCache} to obtain the community (line 8). 
Within the \textit{ComputeHomoWithCache} function, we utilize $size$ to record the traversal count and $\mathcal{P}NbSet$ to record intermediate structures (line 12). Within the loop, if it has been processed of current $v$ and the traversal count for the current iteration of the for loop is equal to length of $S$, we recursively execute the \textit{ComputeHomoWithCache} function with new parameters, or proceed to the next loop iteration (lines 15-18). We then ascertain all $\mathcal{P}$-neighbors of the current vertex $v$, intersect them with $\mathbb{R}$, and update related container (lines 19-23). When the iteration count of the for loop is equal to the length of $S$, we recursively invoke the \textit{ComputeHomoGraph} function with the current parameters (line 25). Ultimately, the function returns the $Adj$ map (line 26). For the remaining adjacency matrix, we proceed significance in the same post-processing manner as in Algorithm~\ref{algo:QHSC} and return the HSC $\mathcal{C}$.

Algorithm~\ref{algo:Advanced} illustrates the method of reusing the solution space based on nested meta-paths. Similarly, when a group of queries contains queries with larger k values under a fixed meta-path, the solution space of smaller $k$ can also be reused. Due to the limitation of space, it is not elaborated here.

\section{Offline Index Method IHSC}
Optimizations to online query algorithms, despite their advancements, still exhibit inherent limitations in handling large-scale graph data in real-world contexts. Thus, employing index-based queries and optimization methods becomes imperative. In response, we introduce the IHSC index structure for the SACH-Problem, this design is grounded onthe following observations or conclusions.

\text{\textbf{Meaningful meta-path is finite.}} In HINs, meta-paths are designed for practical applications rather than just serving as abstract mathematical concepts. Data mining research predominantly focuses on discerning general patterns in network data. Both the heterogeneous graph data and the meta-paths possess distinct characteristics that are integrated into research goals.  Consequently, some theoretically plausible meta-paths may not align with overarching research objectives.
Consider Fig.~\ref{fig:IMDB}: an M-type vertex commonly connects to only one D-type vertex, mirroring to the fact that a movie typically has only one director. Yet, a meta-path $\mathcal{P}$=(\textit{DMD}) can be defined, signifying a quest for two directors associated with the same movie. Clearly, it will yield very few instances of $\mathcal{P}$. 
Furthermore, in research, the length of meta-paths is usually limited. According to some sociological theories, sufficiently long meta-paths might connect vertices in the HIN, they represent very weak relationships. Overall, the number of meta-paths in HINs, as a research subject, is limited.

\vskip 3pt
\begin{lemma}
\label{lemma:result community nesting}
\textbf{The significance between vertices can be transferred to HSCs.} Given an HIN, a meta-path $\mathcal{P}$, and a positive integer $k$, if $v_1$ and $v_2$ are $\mathcal{P}$-connected, and the respective HSC results for $q_1$ and $q_2$ are $\mathcal{C}_1$ and $\mathcal{C}_2$. If $Sig(v_1) \leq Sig(v_2)$ then $\mathcal{C}_2 \subseteq \mathcal{C}_1$.    
\end{lemma}
\vskip 3pt
\begin{proof} From the given conditions, it is evident that $f(\mathcal{C}_1) \leq Sig(v_1)$ and $f(\mathcal{C}_2) \leq Sig(v_2)$. Assuming $\mathcal{C}_1 \subseteq \mathcal{C}_2$ and in light of \textit{Lemma}~\ref{lemma:attribute value relationship}, it follows that $f(\mathcal{C}_1) \geq f(\mathcal{C}_2)$, which contradicts the existing conclusion $Sig(v_1) \geq Sig(v_2)$. Thus, the only feasible inclusion relationship between $\mathcal{C}_1$ and $\mathcal{C}_2$ is $\mathcal{C}_2 \subseteq \mathcal{C}_1$, validating \textit{Lemma}~\ref{lemma:result community nesting}.
\end{proof}
\vskip 3pt

\subsection{IHSC Structure}
Building upon the previously discussed observations and lemmas, we introduce the foundational framework of Index of HSC (IHSC). The following describes the conceptualization and construction:
\begin{enumerate}
    \item Create distinct indices based on the $k$ values for the limited number of meta-paths in the HINs.
    \item Within each index corresponding to a specific value of $k$, organize all vertices that are $\mathcal{P}$-connected to the query vertex and satisfy $\alpha(v, \mathcal{C}) \geq k$ in a tree structure. Each subtree is capable of forming an HSC, with significance defined by the minimum significance value among the root nodes of the subtrees.
\end{enumerate}

\begin{figure}[t]
    \centering
    \includegraphics[width=1\columnwidth]{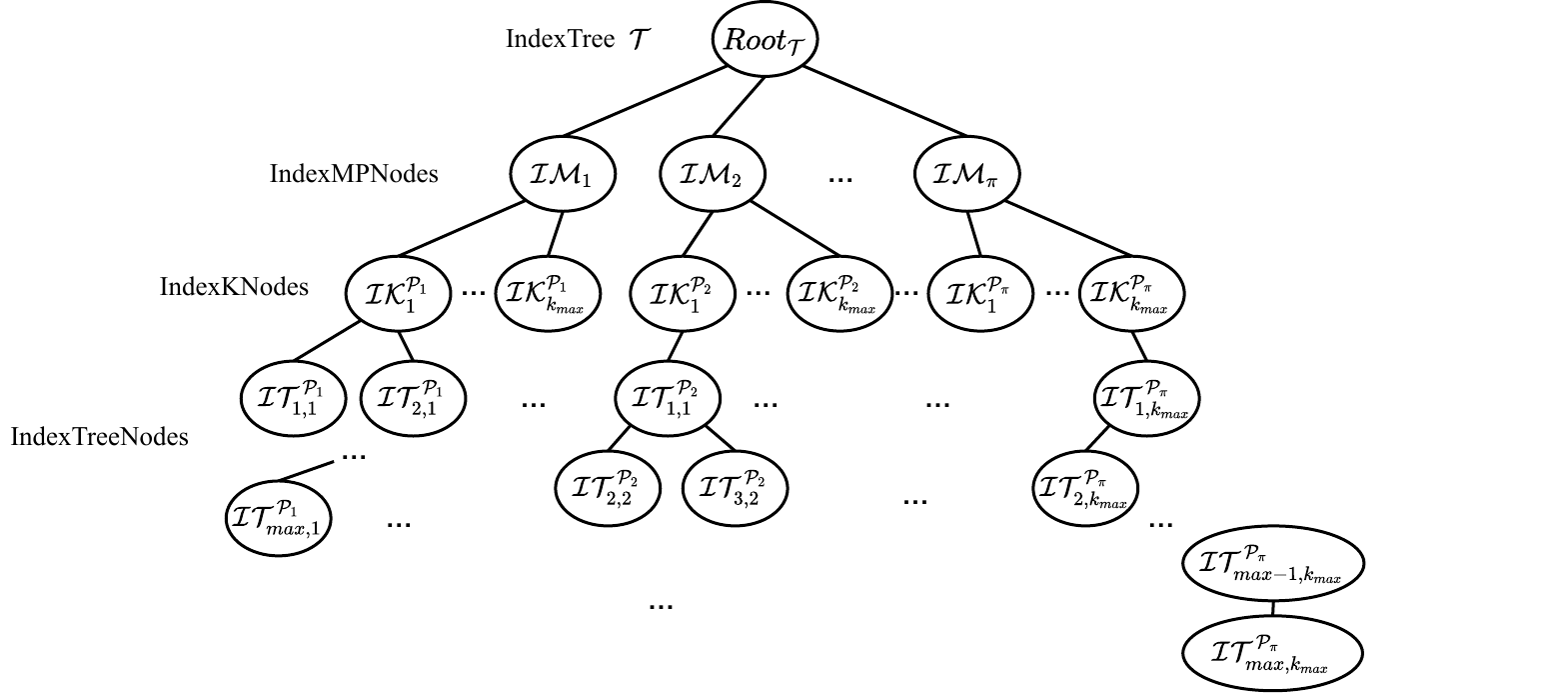}   
    \vspace{-10pt}
    \caption{Basic IHSC structure}
    \vspace{-10pt}
    \label{fig:basic-index}
\end{figure}

Specifically, Fig.~\ref{fig:basic-index} delineates the structure of IHSC. The root node of the tree, denoted as $\mathcal{T}$, functions as the \textit{IndexTree}. 
$\mathcal{T}$ is structured in such a way that it starts from the root node, employing each of the $\Pi$ meta-paths as the primary level index.  We create \textit{IndexMPNodes}, symbolized as $\mathcal{IM}_i$ where $i \in \left[1, \Pi\right]$, for each meta-path. 
Under every $\mathcal{IM}i$, there exist $k_{\text{max}}^{\mathcal{P}i}$ second-level \textit{IndexKNodes} represented  as $\mathcal{IK}_j^{\mathcal{P}i}$ where $j \in \left[1, k_{\text{max}}^{\mathcal{P}_i}\right]$. 
Below is the third level, which consists of the HSC vertices subtrees that correspond to the meta-path $\mathcal{P}_i$ and parameter $j$.
Each HSC forms a subtree consisting of several \textit{IndexTreeNode} denoted as $\mathcal{IT}$, instead of a solitary tree node. A singular $\mathcal{IT}$ node can map to one or multiple vertices. This is because we reflect the characteristics exhibited by significance within the subtree. The significance of a node is depicted by the minimum significance across the vertices within that node. Additionally, the graph vertices stored within each $\mathcal{IT}$ node indicate that these vertices share the same processing priority and must be operated simultaneously, without individually extracting a single vertex from the $\mathcal{IT}$ node. 
Furthermore, the parent-child relationship between $\mathcal{IT}$ nodes reflects the priority displayed by significance, where the significance of the parent $\mathcal{IT}$ node is always smaller than that of its child $\mathcal{IT}$ nodes. In our two indexes, the correspondence between the index node types and symbols used is shown in Table~\ref{tab:treenode}.

\begin{table}
\centering
\caption{Indexes nodes type and symbol comparison}
\label{tab:treenode}
\vspace{-3pt}
\begin{tabular}{|c|c|} 
\hline
\textbf{Type} & \textbf{Symbol}   \\ 
\hline
\textit{IHSC}           & $\mathcal{T}$           \\ 
\hline
\textit{OIHSC}          & $\mathcal{T}_{A}$          \\ 
\hline
\textit{IndexMPNode}    & $\mathcal{IM}$         \\ 
\hline
\textit{IndexKNode}       & $\mathcal{IK}$        \\
\hline
\textit{IndexTreeNode}       & $\mathcal{IT}$        \\
\hline
\textit{Virtual-IndexTreeNode}       & $\mathcal{VIT}$        \\
\hline
\end{tabular}
\end{table}

\subsection{IHSC Constructing Method}

Algorithm~\ref{algo:basic-index} showing the establishment process, first initialize $\mathcal{T}$ and $\mathcal{IM}$s according to the given conditions.
For each $j \in \left[1, kMax\right]$, initialize an $\mathcal{IK}_j$ and set it as a child node of $\mathcal{IM}_i$. Compute all connected components under $j$-core and invoke the \textit{CreateIACHIndexTree} function for each connected component (lines 10-14). 

\begin{algorithm}[t]
\small
\caption{IHSC Constructing}
\label{algo:basic-index}
\begin{algorithmic}[1]
\State \textbf{Input:} HIN $G = (V, E, T, \varphi, \psi)$, a positive integer $k$, several meta-paths $\mathcal{P}_i$ where $1 \leq i$ $\land$ $i \in \mathbb{Z}^+$, query vertex $q$.
\State \textbf{Output:} One IndexTree $\mathcal{T}$ of $G$
\State Initialize IndexTree $\mathcal{T} \gets \varnothing$
\State Initialize several IndexMPNode $\mathcal{IM}_i$ according to $\mathcal{P}_i$
\For{each $\mathcal{IM}_i$}
    \State $\mathcal{T}$.addchild($\mathcal{IM}_i$)
    \State $NeiMap\gets$the adjacency matrix by decomposing $G$
    \State $kMax \gets$ max value of the decomposing result
    \For{each $j \in \left[1, kMax\right]$}
        \State Initialize IndexKNode $\mathcal{IK}_j\gets j, \mathcal{IM}_i$
        \State $\mathcal{IM}_i.addchild(\mathcal{IK}_j)$
        \State $\mathcal{CP}s \gets$ all the $j$-components of $G$
        \For{each $\mathcal{CP} \in \mathcal{CP}s$}
            \State CreateIACHIndexTree($\mathcal{CP}, \mathcal{IK}_j, j, NeiMap$)
        \EndFor
    \EndFor
\EndFor
\State \Return $\mathcal{T}$
\end{algorithmic}
\end{algorithm}

\begin{algorithm}
\small
\caption{CreateIACHIndexTree}
\label{algo:createIACHIndexTree}
\begin{algorithmic}[1]  
\State \textbf{Function} \textit{CreateIACHIndexTree($\mathcal{CP}, \mathcal{IK}, k, Map$)}
\State $Q \gets \varnothing$, $delete \gets \varnothing$, $reserve \gets \varnothing$
\State $minV \gets$ vertices with minimum significance
\State $Q\gets$$Q$.addAll($minV$)
\While{$Q \neq \varnothing$}
    \State $cur\gets Q.poll()$, $nei \gets Map.getValue()$
    \For{each $n \in nei$}
        \State $degree \gets Map$.get($n$).length$-1$
        \If{$degree<k$}
            \State $reserve$.add($cur$)
            \If{$!Q$.contains($n$) $\land !delete$.contains($n$)}
                \State $Q$.add($n$)
            \EndIf
        \EndIf
    \EndFor
    \State $delete$.add($cur$)
    \If {$!Q$.contains($cur$)}
        \State Remove $cur$ and its associated edges from $Map$
        \State $reserve.$remove($cur$)
    \EndIf
\EndWhile
\State Remove vertices from $reserve$ and associated edges from $Map$
\State Initialize IndexTreeNode $\mathcal{IT}$ with $delete$
\State $\mathcal{IK}.addChild(\mathcal{IT})$
\State $Child\mathcal{CP}s\gets$the set of connected components after remove vertices in $delete$
\For{each $Child\mathcal{CP} \in Child\mathcal{CP}s$}
    \State CreateIACHIndexTree($Child\mathcal{CP}, \mathcal{IK}, k, Map)$
\EndFor
\State \Return $\mathcal{IK}$
\end{algorithmic}
\end{algorithm}

The \textit{CreateIACHIndexTree} function, as presented in Algorithm~\ref{algo:createIACHIndexTree}, aims to enhance the mapping correspondence between graph vertices and tree nodes for every $\mathcal{IK}$.
Initially, the vertices with the lowest significance are identified and enqueued into $Q$.
The following operations are performed while $Q$ is not empty: the foremost element in $Q$ is retrieved as $cur$, if its degree minus one is less than $k$, $cur$ is added to $reserve$(lines 7-12). 
Should $cur$ not exist in $Q$, it, along with its affiliated edges, is expunged from the adjacency map and is similarly eliminated from $reserve$ (lines 14-16).
Upon the termination of the while loop, all vertices housed in the reserve collection are deleted from the map (line 17).
Tree node $\mathcal{IT}$ is initialized with every vertex present in $delete$, forging a hierarchical bond with $\mathcal{IK}$ (lines 18-19). 
Given the unpredictability concerning the formation of new sub-connected components post vertex elimination within a $k$-connected cluster, a further assessment and acquisition of these new components are undertaken (line 20).
For each connected component, the \textit{CreateIACHIndexTree} function is recursively called (lines 21-22) and finally returns the $\mathcal{IK}$ node (line 23).

\subsection{IHSC Query Method}
Leveraging the architecture of IHSC, we introduce the associated query method. The tree skeleton has natural advantages in traversal and search, we will make full use of this feature, the query algorithm is carried out as below.

\begin{algorithm}
\small
\caption{Querying IHSC}
\label{algo:Basic index Query}
\begin{algorithmic}[1]  
\State \textbf{Input:} IHSC $\mathcal{T}$, a set of parameters as follows: a positive integer $k$, a meta-path $\mathcal{P}$ , a query vertex $q$ or a query significance value $w$.
\State \textbf{Output:} Corresponding result SACH-Communities
\State$\mathcal{IM}^* \gets \mathcal{IM}$ whose meta-path=$\mathcal{P}$ 
\State$\mathcal{IK}^* \gets \mathcal{IK}$ whose $core_{value}=k$  
\State FindFitNodes($\mathcal{IK}^*$, $q$ \textbf{or} $w$)
\Statex
\State \textbf{Function} \textit{FindFitNodes($node$, $query$)}
\State $\mathcal{C} \gets \varnothing$
\If{$node.type=\mathcal{IK}$}
    \For{each $\mathcal{IT} \in node.children$}
        \State $\mathcal{C}.$addAll($FindFitNodes$($\mathcal{IT}, query$))
    \EndFor
\EndIf
\If{$node.type=\mathcal{IT}$}
    \If{Given $q$}
        \If{$node.vertices.$contains($query$)}
            \State $\mathcal{C}$.add($node$)
        \Else
            \For{each $\mathcal{IT} \in node.children$}
                \State $\mathcal{C}.$addAll($FindFitNodes(\mathcal{IT}, query$))
            \EndFor
        \EndIf
    \EndIf
    \If{Given $w$}
        \If{$node.minValue \geq w$}
            \State $\mathcal{C}$.add($node$)
        \Else
            \For{each $\mathcal{IT} \in node.children$}
                \State $\mathcal{C}.$addAll($FindFitNodes(\mathcal{IT}, query$))
            \EndFor
        \EndIf    
    \EndIf  
\EndIf
\State \Return $\mathcal{C}$
\end{algorithmic}
\end{algorithm}

In Algorithm \ref{algo:Basic index Query}, we first obtain the corresponding $\mathcal{IM}^*$ and $\mathcal{IK}^*$ using the tree traversal matching method based on the input meta-path $\mathcal{P}$ and positive integer $k$. 
Then, the \textit{FindFitNodes} function is called to retrieve the root nodes and their subtrees corresponding to the communities, which is a function that can handle multiple types of tree nodes. 
The search procedure involves a top-down traversal of vertices and significance starting from $\mathcal{IK}^*$.
When inputting vertex $q$, find $\mathcal{IT}$ containing $q$; When the input significance $w$, find the first node $\mathcal{IT}$ whose significance is not less than $w$. 
Upon pinpointing the suitable node, denoted as $\mathcal{IT}^*$, the vertex set from this node and all its descendant nodes is used to produce the resulting HSC.

\begin{figure}[t]
    \centering
    \hspace*{0.5cm}
    \includegraphics[width=0.9\columnwidth]{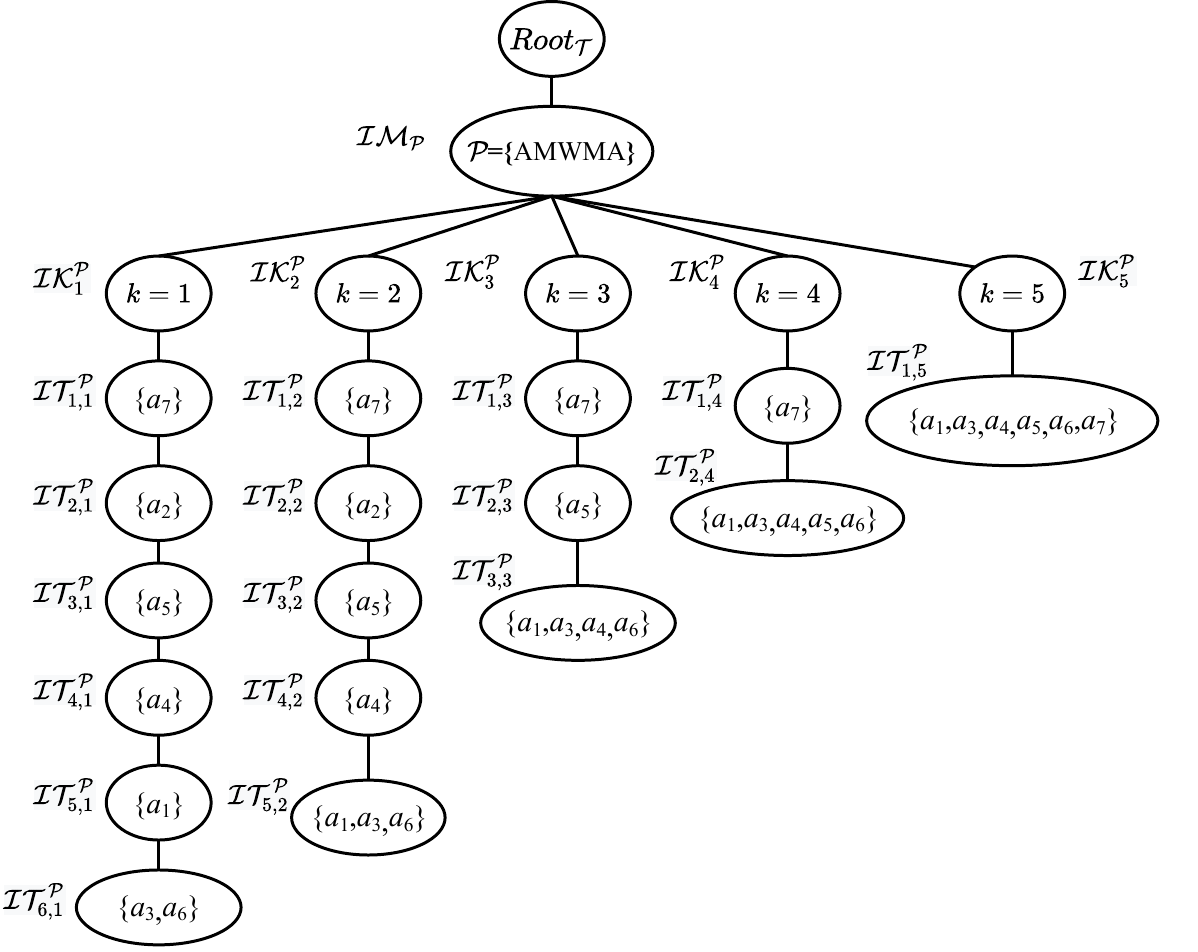}   
    \vspace{-3pt}
    \caption{An example of IHSC}
    \vspace{-15pt}
    \label{fig:building basic-index}
\end{figure}

\begin{example}
Consider the HIN shown in Fig.~\ref{fig:IMDB} with A-type as the anchor-type and meta-path $\mathcal{P}$ = (\textit{AMWMA}), we commence index construction from $k$ = 1. 
Initially, the root node \( \mathcal{R}oot_\mathcal{T} \) of the index tree \( \mathcal{T} \) and \( \mathcal{IM}_{\mathcal{P}} \) are established. 
Subsequently, \( \mathcal{IM}_{\mathcal{P}} \) is linked as a child node under \( \mathcal{R}oot_\mathcal{T} \).
Analyzing \( \mathcal{P} \) connectivity of graph \( G \), it's discerned that \( k_{max} \) = 5. Accordingly, nodes \( \mathcal{IK}_1 \) through \( \mathcal{IK}_5 \) are constructed for \( k \) within the range [1,5] and associated as child nodes of \( \mathcal{IM}_{\mathcal{P}} \). In this instance, for every \( k \) in the range [1,5], a singular connected component is created.
For \( k \) values 1 and 2, all A-type vertices adhere to the 1-core and 2-core structural prerequisites. The community hierarchy can thus be derived from the significance values. When \( k \geq 3 \), vertex \( a_2 \) no longer aligns with the structural stipulations, leading to its exclusion from the community. Subsequent tree construction for the remaining vertices is determined by significance constraints. This resultant structure can be viewed in Fig.~\ref{fig:building basic-index}. 
For instance, with \( k \) = 4 and \( q \) = \( a_1 \), the combined vertex sets of \( \mathcal{IT}_{1,1}^{\mathcal{P}} \) and \( \mathcal{IT}_{2,1}^{\mathcal{P}} \) constitute a 4-core community \( \mathcal{H}_1 \) containing vertices \( a_1, a_3, a_4, a_5, a_6, \) and \( a_7 \) with \( f(\mathcal{H}_1) \) = 1. Given that the query vertex $q\neq a_7$, which is not mandatory, the community \( \mathcal{H}_2 \), derived solely from the vertex set of \( \mathcal{IT}_2^{\mathcal{P}} \), meets the HSC criteria with \( f(\mathcal{H}_2) \) = 3. Consequently, \( \mathcal{C}(\mathcal{P},4,a_1) \) equals \( \mathcal{H}_2 \). By analogous logic, \( \mathcal{C}(\mathcal{P},4,a_7) \) equates to \( \mathcal{H}_1 \).
\end{example}

\section{Optimal Index Method OIHSC}

\subsection{OIHSC Structure}
The inherent simplicity and coherence of IHSC's design are evident. However, its tendency for redundancy when archiving HSCs results in notable spatial overheads. To address this inefficiency, we introduce the Optimal IHSC (OIHSC) approach.

In the OIHSC framework, it is specifically designed to minimize redundant storage. 
Leveraging \textit{Lemma}~\ref{graph nesting}, we initially employ the solution space of parent meta-paths to expedite the construction of the sub-meta-path's. 
Furthermore, given the inherent nesting characteristics of the $k$-core, we present the node structure under the subtree $\mathcal{IK}_{k_{max}}$.
For preceding trees, when vertices are stored within a tree node, we incorporate an auxiliary storage detailing the its depth and capacity mapping for the vertex. We then store the entire set of vertices in the first ancestor node without sibling relative to the current node. 
At a global level, each $\mathcal{IK}$ logs the nodes genuinely archived under its subtree while preserving a universal mapping from the graph vertices to the subtree of $\mathcal{IK}$.

\begin{figure}[t]
    \centering
    \includegraphics[width=0.95\columnwidth]{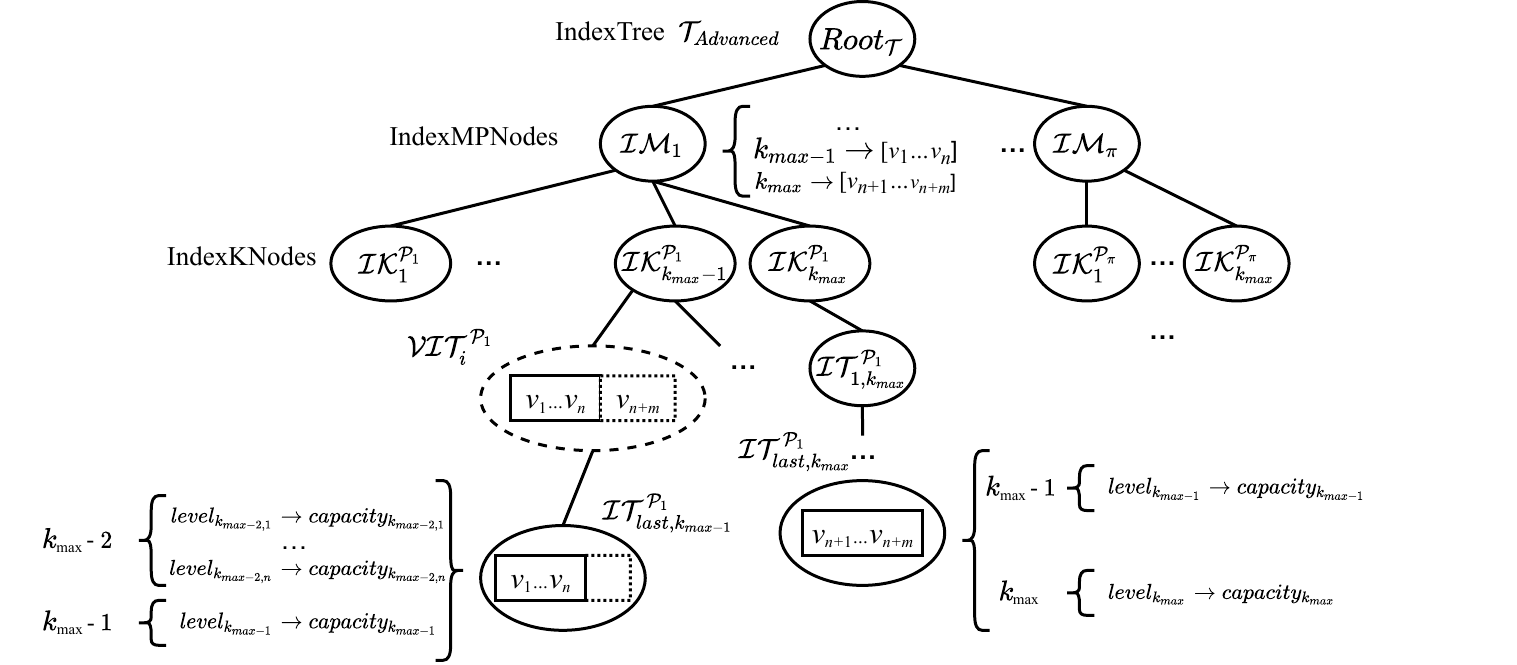}
    \vspace{-8pt}
    \caption{OIHSC structure}
    \vspace{-15pt}
    \label{fig:advanced index structure}
\end{figure}

Fig.~\ref{fig:advanced index structure} depicts the structure, mirroring the multi-level framework of IHSC, with the distinction that each graph vertex is stored only once. 
Inspecting from the bottom upwards, we identify a node, $\mathcal{IT}_{last,k_{max-1}}^{\mathcal{P}_1}$, beneath $\mathcal{IK}_{k_{max}-1}^{\mathcal{P}_1}$, encompassing graph vertices ${v_1, \dots ,v_n,v{n+1},\ldots,v_{n+m}}$. These vertices contribute to forming the ($k_{max}$-1)-core within the homogeneous graph derived from meta-path $\mathcal{P}_1$. Only the vertices ${v_1, \dots ,v_n}$ are stored at this node, while vertices ${v_{n+1},\ldots,v_{n+m}}$ are allocated in the subsequent subtree of $\mathcal{IK}_{k_{max}}^{\mathcal{P}_1}$.
Simultaneously, a \textit{virtual IndexTreeNode} $\mathcal{VIT}_{i}^{\mathcal{P}_1}$ is instantiated at the proximate ancestor node with sibling nodes of the $\mathcal{IT}_{last,k_{max-1}}^{\mathcal{P}_1}$ node. This signifies the $i$-th \textit{virtual IndexTreeNode} under the subtree of  $\mathcal{IK}_{k_{max}-1}^{\mathcal{P}_1}$, storing the aggregate of graph vertices from this node downward.
Further traversal upwards to the $\mathcal{IM}$ node reveals a globally maintained mapping, correlating the $k$ values with the precisely stored graph vertices under the meta-path $\mathcal{P}_1$.

Based on the aforementioned IHSC structure, it exhibits certain properties that can be harnessed to optimize storage space during index construction. We propose the following lemma:
\vskip 3pt
\begin{lemma}
For any of a graph vertex $v$ and an IHSC tree $\mathcal{T}$, for any two adjacent $\mathcal{IK}_{k-1}$ and subtrees under $\mathcal{IK}_{k}$, there are: 
\begin{equation}
\label{eq2}
    depth(v,\mathcal{IK}_k) \leq depth(v,\mathcal{IK}_{k-1})
\end{equation}
The inequality means that the depth of $\mathcal{IT}_k$ in which $v$ is stored under the $\mathcal{IK}_k$ node is necessarily no greater than its depth of $\mathcal{IT}_{k-1}$ under $\mathcal{IK}_{k-1}$. Here we define the depth of all the $\mathcal{IK}$ nodes as 0 and its directing children as 1.
\end{lemma}
\vskip 3pt

\subsection{OIHSC Constructing Method}

\begin{algorithm}
\small
\caption{OIHSC Constructing}
\label{algo:advanced index}
\begin{algorithmic}[1]
\State \textbf{Input:} Skeleton created $\mathcal{T}$ in which $\mathcal{IM}, \mathcal{IK}, \mathcal{IT}$ nodes are created and processed, and link relationship has been established
\State \textbf{Output:} Space saving advanced index $\mathcal{T}_{A}$
\For{each $\mathcal{IK}_{i-1}, \mathcal{IK}_i \in \mathcal{T}$}
    \State $\mathcal{T}_i \gets$ 
    \textit{processKNodes($\mathcal{IK}_{i-1},\mathcal{IK}_i, \mathcal{T}_{i-1}$)}
\EndFor
\State \Return $\mathcal{T}_{A}$
\Statex
\State \textbf{Function} \textit{processKNodes($\mathcal{IK}_{i-1},\mathcal{IK}_{i}, \mathcal{T}_{i-1}$)}
\State $Actual \gets \varnothing$
\State $S_1 \gets$ all descendant $\mathcal{IT}$ of $\mathcal{IK}_{i-1}$
\State $S_2 \gets$ all descendant $\mathcal{IT}$ of $\mathcal{IK}_{i}$

\State $Leaves \gets$ all leaf nodes of $\mathcal{IK}_{i-1}$, $Leader \gets \varnothing$
\State $Leader \gets$ all the leader nodes of each leaf in $Leaves$

\State $V \gets$ all vertices of $S_2$
\For{each $s \in S_2$}
    \For{each $\mathcal{IT} \in \mathcal{IK}_{i-1}.children$}
        \If{$\mathcal{IT}.depth \geq s.depth$}
            \State $Node_s.$add($\mathcal{IT}$)
        \EndIf
    \EndFor
    \For{each $item \in Node_s$}
        \If{$item.vertices \cap s.vertices \neq \varnothing$}
            \State $DV[item.depth] \gets s.vertices.length$ 
            \State $s.map[\mathcal{IK}_{i-1}.k_{value}] \gets DV$
        \EndIf
    \EndFor                                                                     
\EndFor

\For{each $s \in S_1$}
    \If{$s.vertices \cap V \neq \varnothing$}
        \State $s.vertices$.remove($V$)
        \State $Actual \gets $ the rest vertices 
        \If{$Leader.$contains(s)}
            \State $s \gets$ communities ruled by $s$
            \State $s.type \rightarrow \mathcal{VIT}$
        \EndIf
    \EndIf
\EndFor
\State $\mathcal{T}_{i-1}.actual \gets Actual$
\State \Return $\mathcal{T}_{i-1}$
\end{algorithmic}
\end{algorithm}
The establishment of OIHSC leverages the existing IHSC $\mathcal{T}$, 
Algorithm \ref{algo:advanced index} details the optimization, processing each subtree under $\mathcal{IK}_k$ for $k \in \left[1,k_{max}\right]$ in function \textit{processKNodes}. 
Each invocation of the \textit{processKNodes} function processes two adjacent subtrees. It sequentially records all descendants and leaf nodes within the subtree, identifying its leader node based on the leaf nodes. Here, the leader node is defined as the earliest ancestor node that possesses sibling nodes.
For each graph vertex under $\mathcal{IK}_{i}$, stored in $S_2$, we augment the $Nodes$ list with nodes from the subtree under $\mathcal{IK}_{i-1}$ that have a depth not inferior to elements in $S_2$. 
Subsequently, for every node $s \in S_2$, if there is an overlap between the vertices of the node elements in $Nodes$ and those of $s$, we log a key-value pairing of \{${depth \rightarrow s.vertices.length}$\} in the internal map $DV$.  
This mapping denotes the correlation between the depth of a tree node and the number of graph vertices it contains (lines 13-20).
For each node $s$ in $S_1$, if there exists an intersection between the vertex set of $s$ and the set $V$, we exclude the intersecting vertices from $s$'s vertex set, documenting the residual vertices. 
For a leader node $s$, we assign a community significance and designate its type as the virtual-indexTreeNode, denoted $\mathcal{VIT}$ (lines 21-27).
Conclusively, we record the vertices contributing the most to the $(i-1)$-core as a mapping on $\mathcal{IK}_{i-1}$, resulting the optimized subtree $\mathcal{T}_{i-1}$.

\subsection{OIHSC Query Method}
Upon deriving the OIHSC, index-based queries involve locating the appropriate subtree and retrieving vertices not stored within that subtree via mapping— a process that's both straightforward and efficient. 
To elucidate, we directly provide a example of the query process.
\begin{algorithm}
\small
\caption{Querying OIHSC}
\label{algo:advanced index query}
\begin{algorithmic}
\State \textbf{Input:} An OIHSC tree $\mathcal{T}_{A}$, a meta-path $\mathcal{P}$, a query vertex $q$ and a positive integer $k$
\State \textbf{Output:} An HSC $\mathcal{C}$

\State $map \gets$ mapping relations stored on $\mathcal{IM}_\mathcal{P}$
\State $\hat{k} \gets map.$get($q$)
\If{$\hat{k} < k$}
    \State \Return $\varnothing$
\EndIf
\If{$\hat{k}$ = $k$}
    \State ${\mathcal{VIT}}^* \gets \mathcal{VIT}$ that $\mathcal{VIT}.vertices$.contains($q$)
    \If{$q \in \mathcal{IT}.vertices$ $\cap$ $\mathcal{IT} \in \mathcal{VIT}^*.children$}
    \State \Return $\mathcal{IT}.vertices$
    \EndIf
\Else
    \State $\mathcal{IT}^{*} \gets$ actual $\mathcal{IT}$ storing $q$ under $\mathcal{IK}_{\hat{k}}$ 
    \While{$\hat{k} \neq k$}
        \State $strMap \gets \mathcal{IT}.map$ that $q \in \mathcal{IT}.vertices$
        \State $\hat{k} \gets \hat{k} - 1$
        \State $List(\mathcal{IT}) \gets strMap.$get($k-1$)
        \State $\mathcal{IT} \gets$ filter by significance from $List(\mathcal{IT})$
        \If{$\hat{k}$=$k$}
            \If{$\mathcal{IT}.vertices \neq \varnothing$}
                \State $\mathcal{VIT}^* \gets$ lowest virtual ancestor of $\mathcal{IT}$ 
                \State $\mathcal{C} \gets$significance fit vertices from $\mathcal{VIT}^*$        
            \Else
                \State $\mathcal{C} \gets k+1$ vertices from $\mathcal{IT}^*.vertices$
            \EndIf
            \State \Return $\mathcal{C}$
        \EndIf
    \EndWhile
\EndIf
\end{algorithmic}
\end{algorithm}

\begin{figure}[t]
    \centering
    \includegraphics[width=\columnwidth]{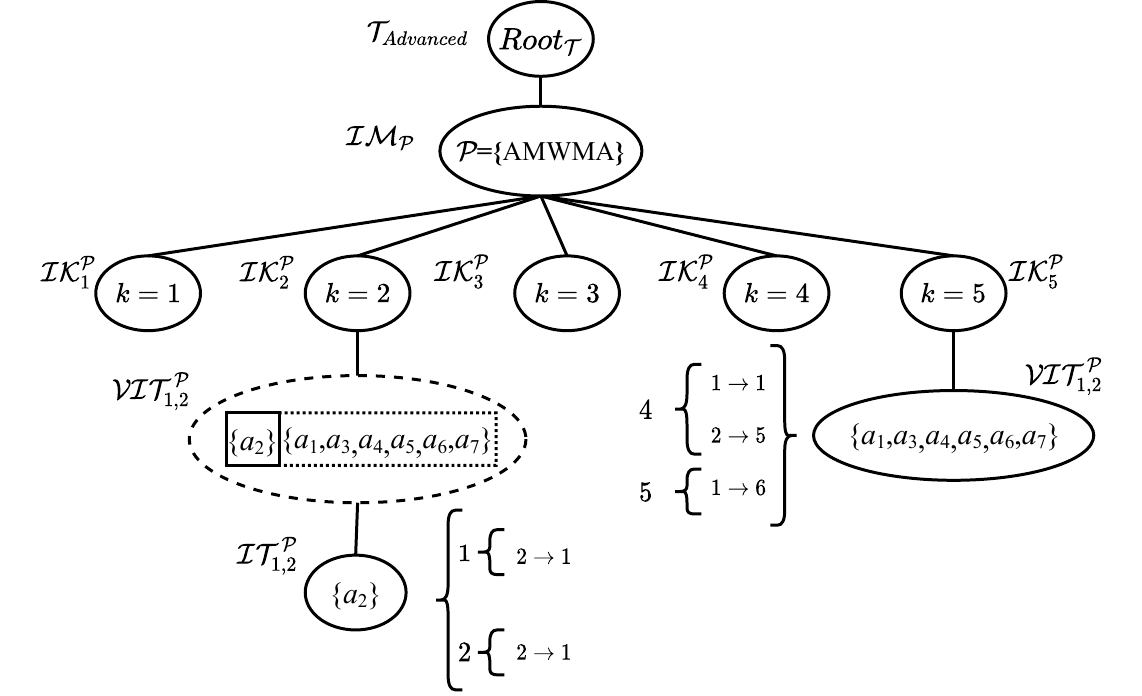}
    \vspace{-18pt}
    \caption{An example of OIHSC}
    \vspace{-11pt}
    \label{fig:advanced index example}
\end{figure}

\begin{example}
 Using the index depicted in Fig.~\ref{fig:advanced index example}, let's consider a query with $k$=4 and $q$=$a_1$. 
By querying the $map$, we discern that $a_1$ is effectively stored in the $k$=5 subtree.  
We navigate to node $\mathcal{IT}_{1,5}^\mathcal{P}$, extracting the node location relationship stored therein. This reveals that within the $k$=4 subtree, the six vertices $\{a_1,a_3,a_4,a_5,a_6,a_7\}$ are encompassed by two $\mathcal{IT}$ nodes in a parent-child configuration. 
These nodes represent one and five graph vertices, respectively. As per the foundational index tree, the vertex $a_7$, which has the minimum significance, is the lone element in the parent $\mathcal{IT}$ node's vertex set. The vertex set of the subsequent $\mathcal{IT}$ node, $\{a_1,a_3,a_4,a_5,a_6\}$, represents the HSC we seek.
   
\end{example}

\section{Experiments}

In the experimental section, we conducted tests on time efficiency by changing meta-paths and the value of $k$; tests on space efficiency by changing the value of $k$ for indexing; tests on the changing trend of significance for result HSCs by altering the dataset size and the value of $k$; and tests on algorithm scalability by varying the dataset size. All algorithms are implemented in Java, and all experiments are conducted on an Ubuntu server with 2.40GHz Intel(R) Xeon(R) Gold 6240R CPU and 512GB memory. The specific configuration is as follows.

\textbf{Datasets.} We use four real-world heterogeneous information network datasets, PubMed, IMDB, DBLP, FourSquare\cite{lu2011pubmed, maas2011learning, ley2009dblp, cho2011friendship}, in our experiments. Based on our data requirements for the research objects in the research process, we filtered out some vertices and edges. Table~\ref{tab:datasets} summarizes the statistics of datasets and basic parameters that we use, of which $N(V_{type})$, $N(E_{type})$ and $N(\mathcal{P})$ denote number of vertex types, of edge types and of meta-path types, respectively. Note that numerical significance are not contained in these HIN datasets, we employ a widely used method in \cite{borzsony2001skyline} to generate numerical significance on each vertices, i.e., \emph{independence}, \emph{correlation} and \emph{anti-correlation}. Due to space limit, we report the results obtained from datasets with independent.

\textbf{Parameters.} We vary  parameters: query vertex $Q$, query significance $W$, structural cohesiveness $k$, meta-path $\mathcal{P}$. Table \ref{tab:parameters} shows the range of parameters and their default values (in bold). In each query, we randomly select 100 vertices, and use their total time as the measure of time efficiency.

    

\begin{figure*}[t]
\centering
\subfigcapskip=-4pt
\subfigure[IMDB(\textit{AMA})] {\label{fig:AMA}
\includegraphics[width=0.216\textwidth]{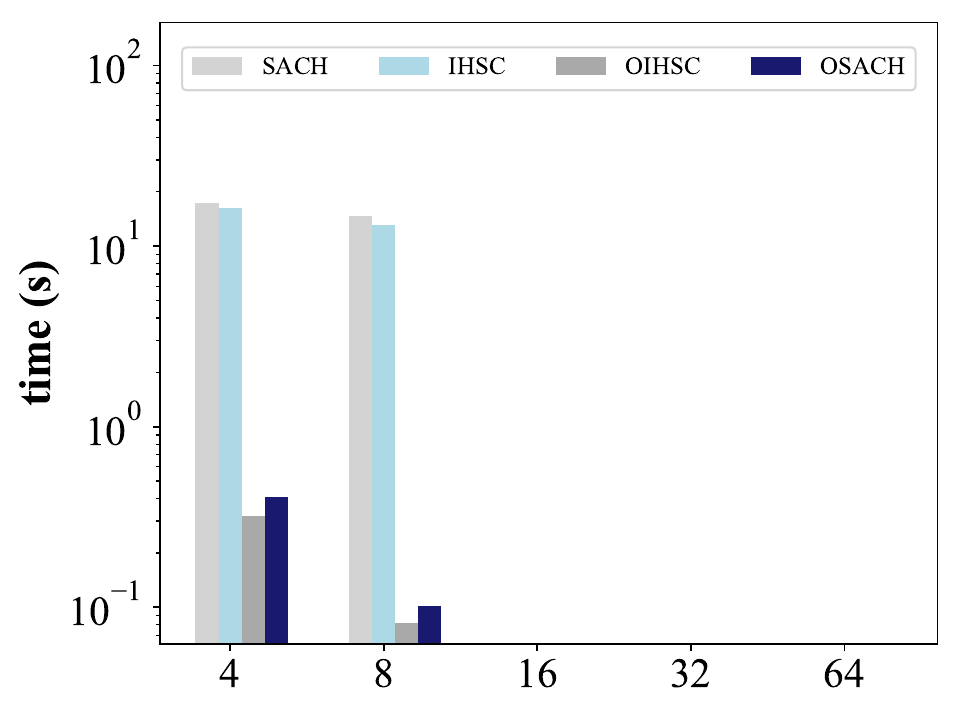}
}
\subfigure[DBLP(\textit{APA})] {\label{fig:APA-t}
\includegraphics[width=0.216\textwidth]{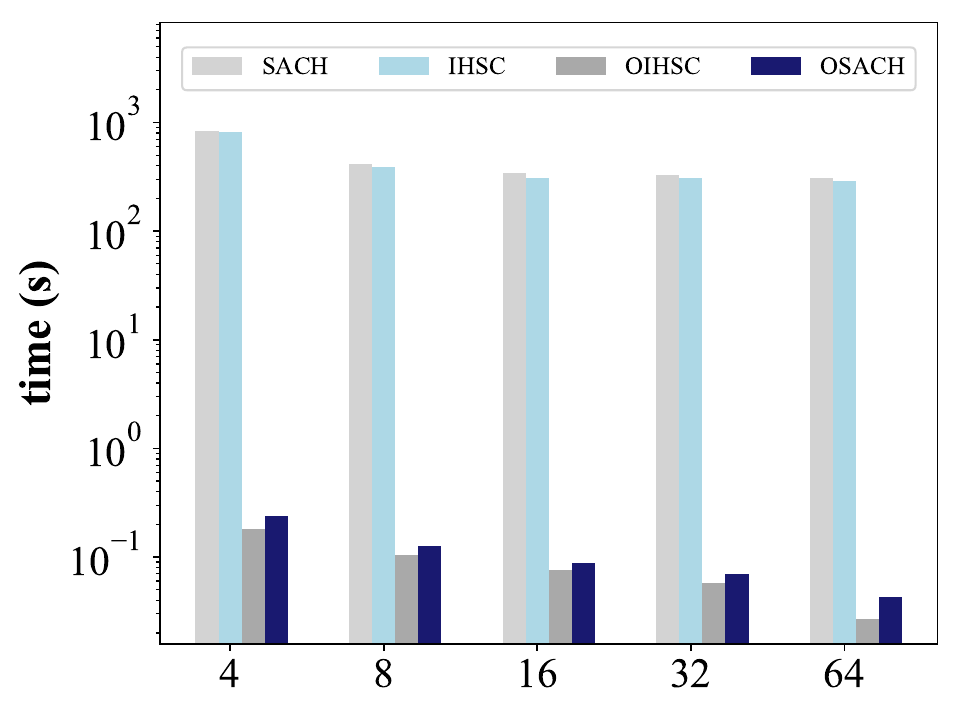}
}
\subfigure[FourSquare(\textit{RVR})] {\label{fig:RVR}
\includegraphics[width=0.216\textwidth]{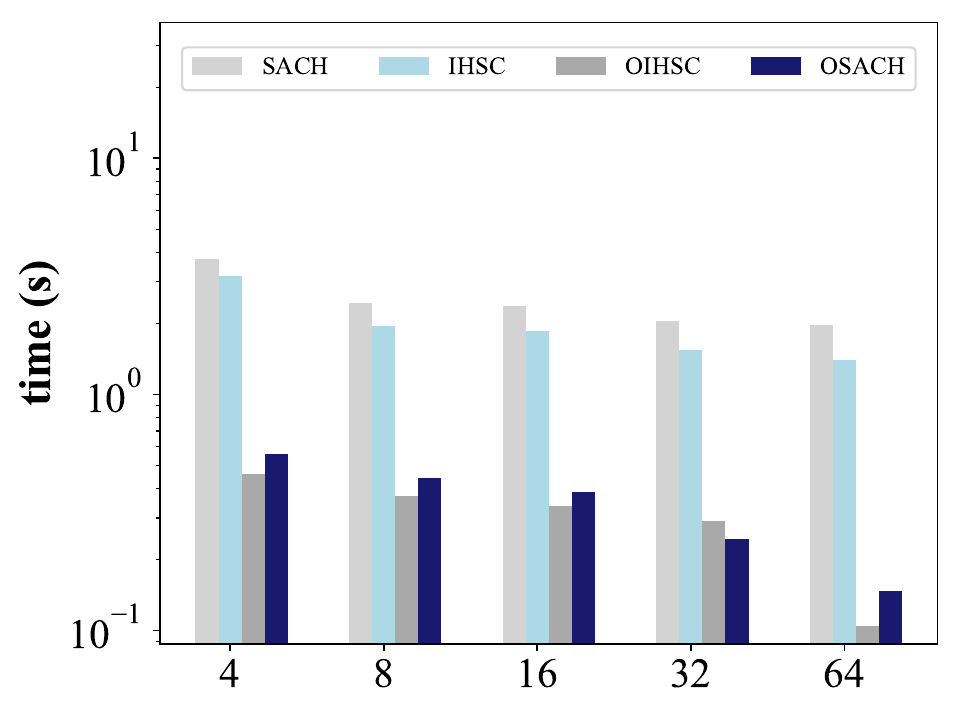}
}
\subfigure[PubMed(\textit{GCG})] {\label{fig:GCG}
\includegraphics[width=0.216\textwidth]{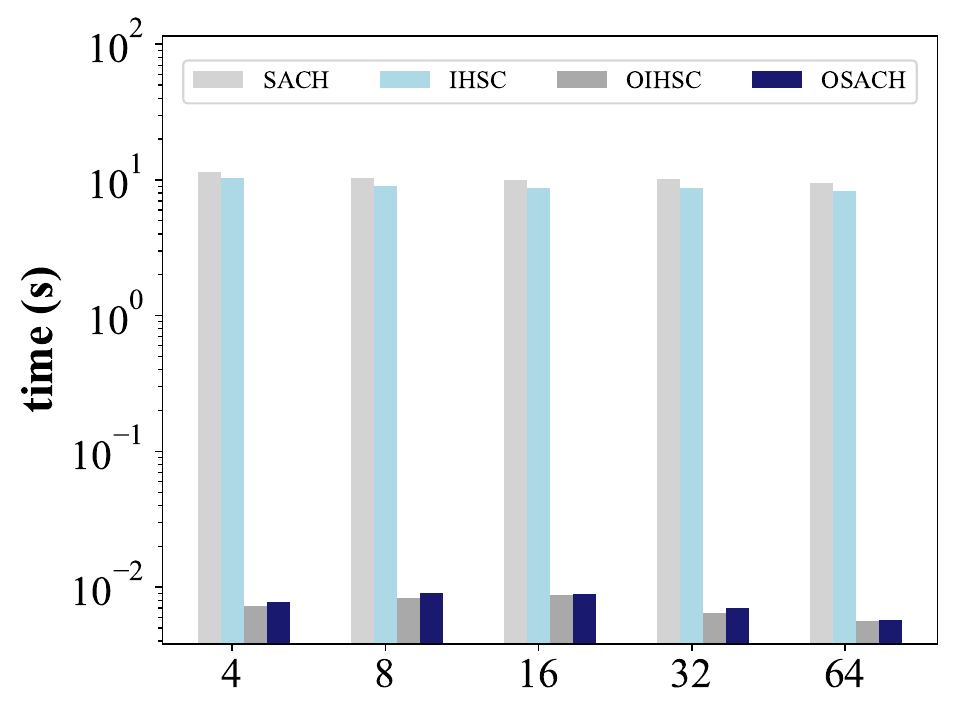}
}
\subfigure[IMDB(\textit{AMWMA})] {\label{fig:AMWMA}
\includegraphics[width=0.216\textwidth]{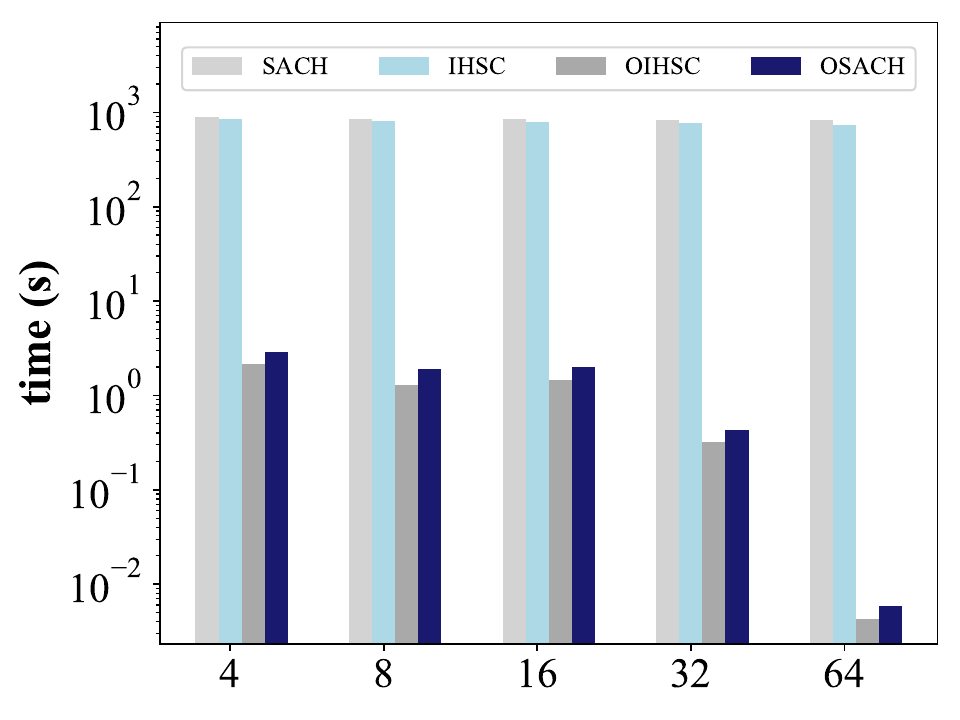}
}
\subfigure[DBLP(\textit{APTPA})] {\label{fig:APTPA}
\includegraphics[width=0.216\textwidth]{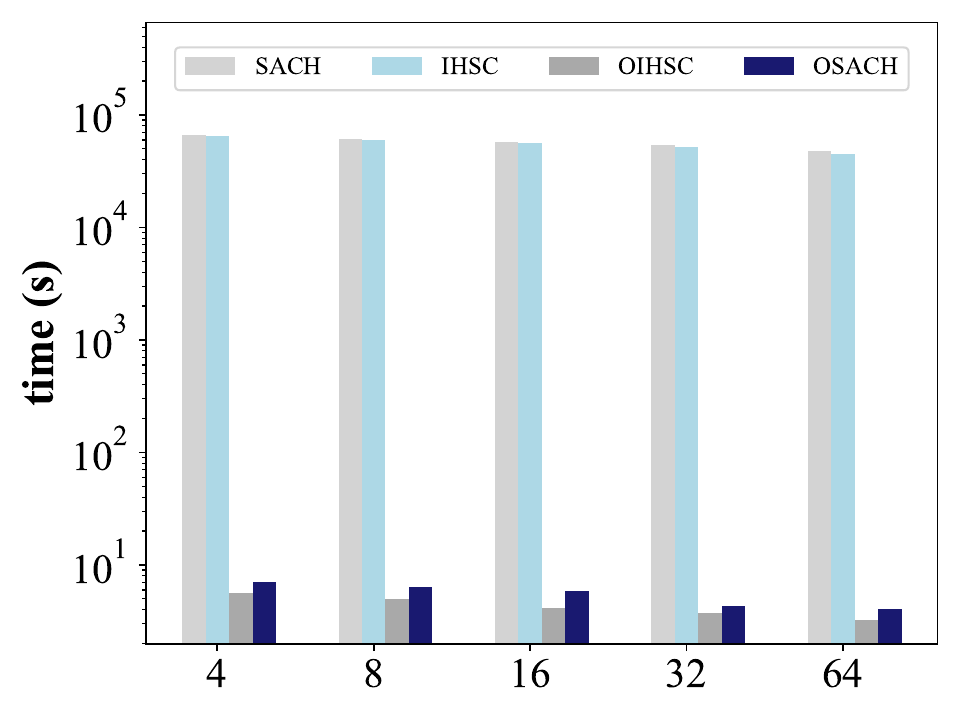}
}
\subfigure[FourSquare(\textit{URVRU})] {\label{fig:URVRU}
\includegraphics[width=0.216\textwidth]{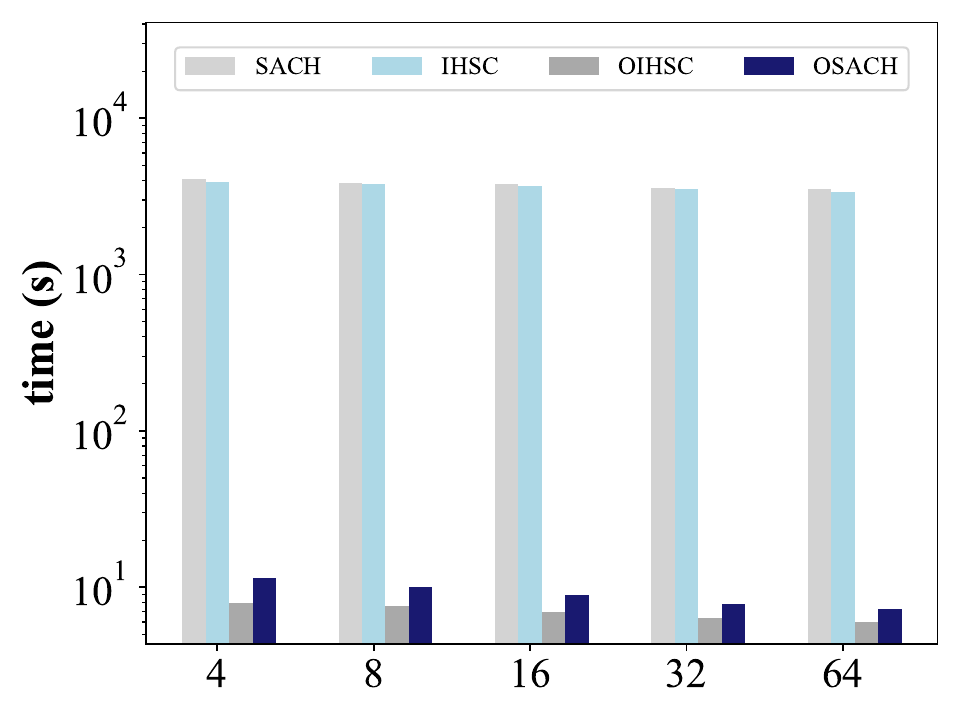}
}
\subfigure[PubMed(\textit{GSG})] {\label{fig:GSG}
\includegraphics[width=0.216\textwidth]{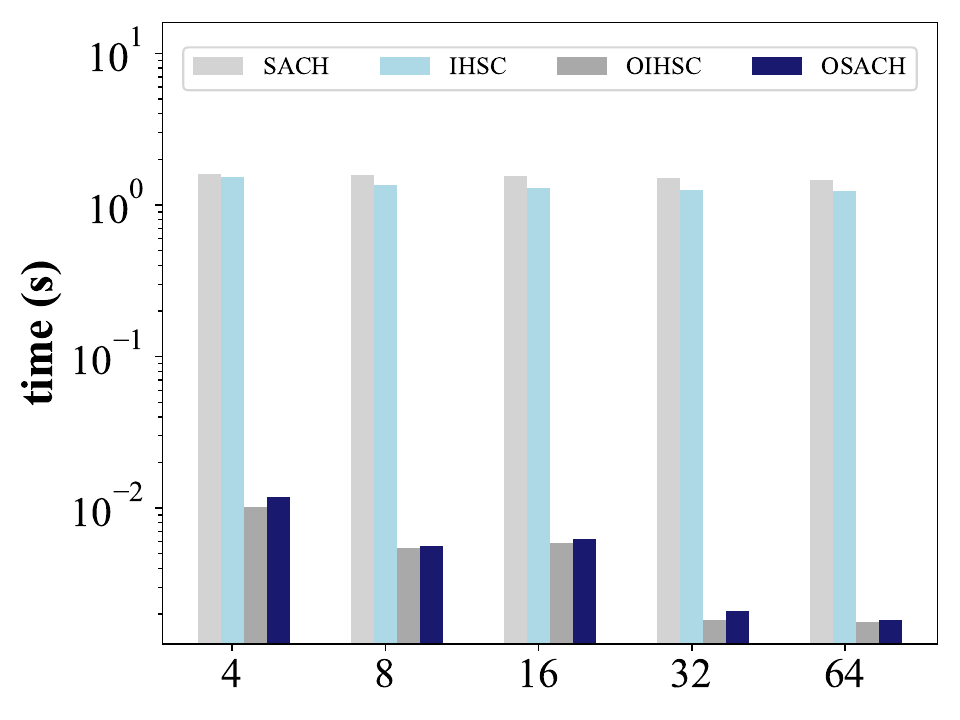}
}
\vspace{-3pt}
\caption{Time efficiency of proposed algorithms in 4 datasets with same meta-path.}
\label{fig:efficiency varingK}
\end{figure*}

\begin{figure*}[t]
\centering
\subfigcapskip=-4pt
\subfigure[PubMed] {\label{fig:PubSpace}
\includegraphics[width=0.216\textwidth]{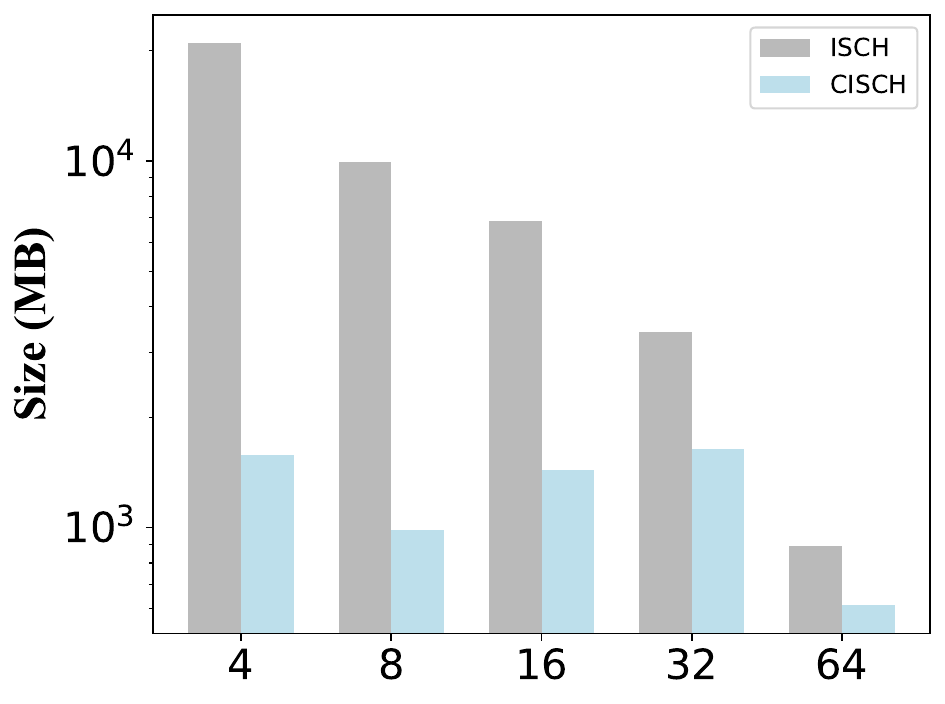}
}
\subfigure[IMDB] {\label{fig:IMDBSpace}
\includegraphics[width=0.216\textwidth]{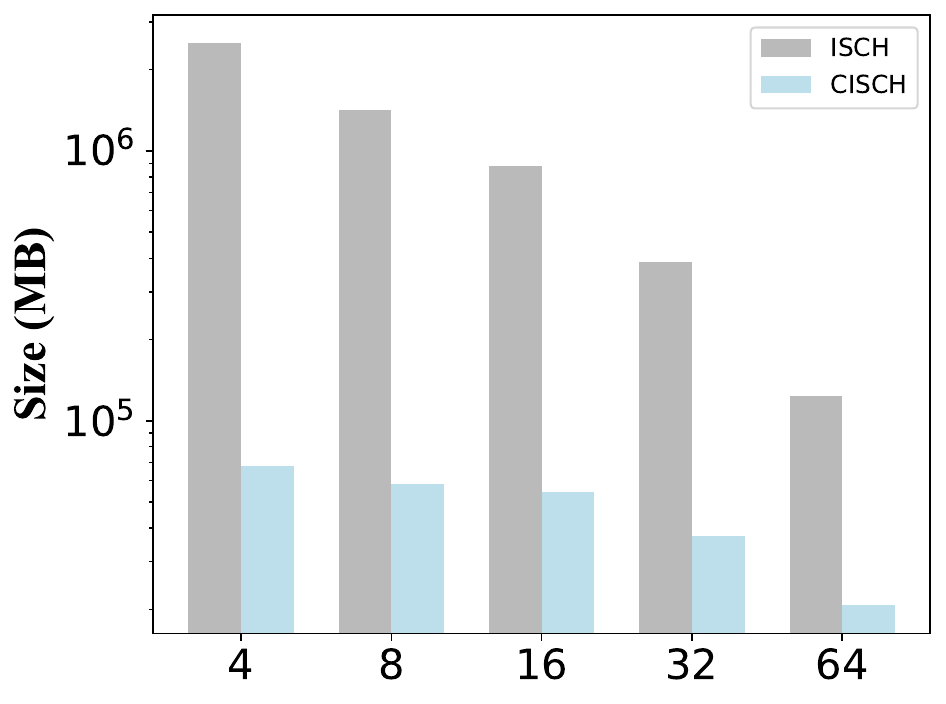}
}
\subfigure[DBLP] {\label{fig:DBLPSpace}
\includegraphics[width=0.216\textwidth]{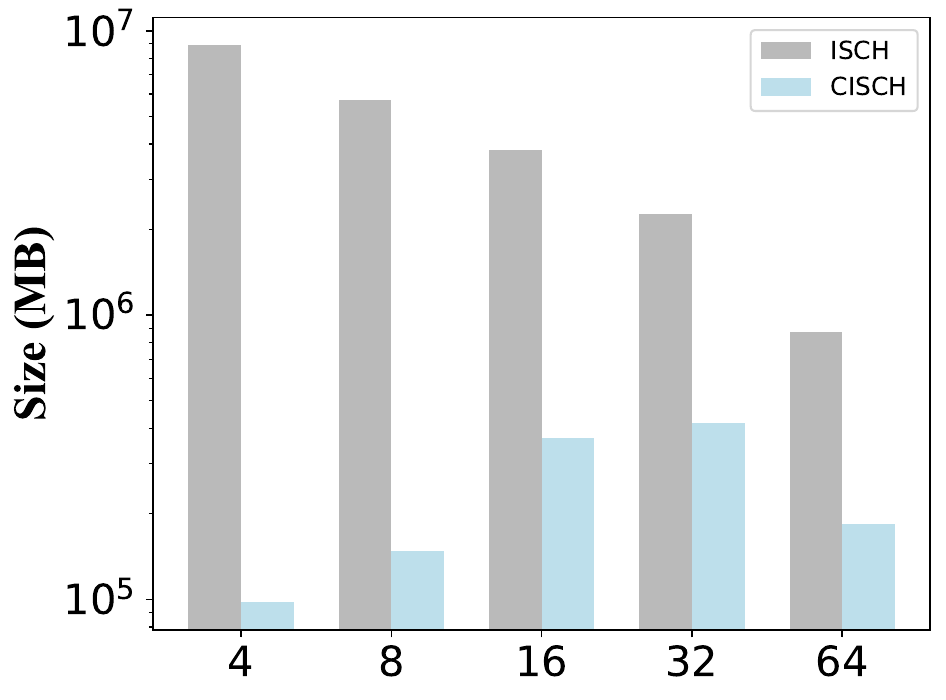}
}
\subfigure[FourSquare] {\label{fig:FSQSoace}
\includegraphics[width=0.216\textwidth]{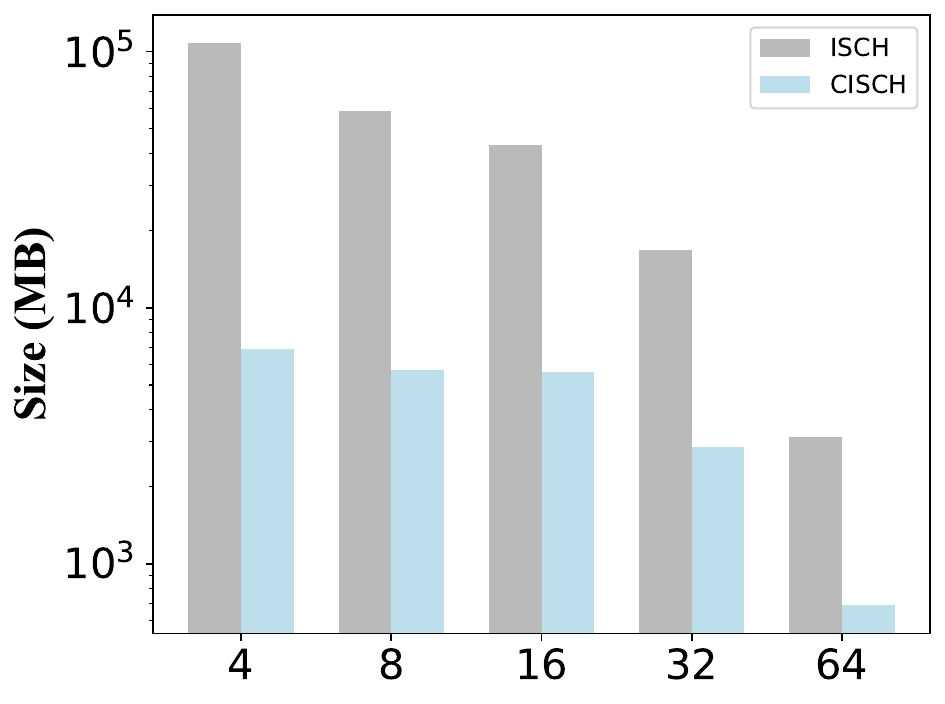}
}
\vspace{-7pt}
\caption{Space efficiency of IHSC and OIHSC in 4 datasets.}
\label{fig:efficiency space}
\vspace{-8pt}
\end{figure*}

\begin{table}
\centering
\caption{Datasets used in our experiments}
\label{tab:datasets}
\vspace{-3pt}
\begin{tabular}{|c|c|c|c|c|c|} 
\hline
\textbf{Dataset} & \textbf{Vertices} & \textbf{Edges} & $N(V_type)$ & $N(E_type)$ & $N(\mathcal{P})$  \\ 
\hline
PubMed           & 14K               & 67K            & 4           & 3           & 6                 \\ 
\hline
IMDB             & 854K              & 7.79M          & 4           & 3           & 10                \\ 
\hline
DBLP             & 2.05M             & 13.2M          & 4           & 3           & 10                \\ 
\hline
FourSquare       & 4.47M             & 20M            & 4           & 3           & 5                 \\
\hline
\end{tabular}
\end{table}

\begin{table}
\centering
\caption{Parameters settings}
\vspace{-3pt}
\label{tab:parameters}
\begin{tabular}{|c|c|} 
\hline
\textbf{Parameters} & \textbf{Tested values} \\ 
\hline
$k$           & 4,8,\textbf{16},32,64  \\    
\hline
{Scale}  & 20\%, 40\%, 60\%, 80\%, \textbf{100\%} \\
\hline
$\mathcal{P}$    &  \makecell[c]{\{\textbf{\textit{RVR}},\textit{URVRU}\},
\{\textbf{\textit{GCG}},\textit{GSG}\},\\
\{\textit{AMA},\textbf{\textit{AMWMA}}\},
\{\textit{APA},\textbf{\textit{APTPA}}\},
}\\ 

\hline
\end{tabular}
\end{table}  

\subsection{Efficiency evaluation}
\textbf{Time efficiency evaluation.} We study the variation in query time under each datasets by altering the value of $k$ and fixing the remaining variables. As shown in Fig.~\ref{fig:efficiency varingK}, we conducted the experiment while maintaining the meta-path $\mathcal{P}$ and the set of index nodes $V$ constant, following the parameters specified in Table~\ref{tab:parameters}. On each dataset, we tested the four query algorithms proposed in this study with three different meta-paths. Taking the IMDB dataset as an example, the results of the 4 community search methods with (\textit{AMA}), (\textit{AMWMA}) as meta-paths are presented in Fig.~\ref{fig:AMA} and Fig.\ref{fig:AMWMA}. It can be observed that Algorithm~\ref{algo:Advanced}, based on meta-path segmentation extension and solution space reuse, performs slightly better than Algorithm~\ref{algo:QHSC}, which simply performs meta-path segmentation extension. Moreover, compared to the online algorithms, the index query Algorithm~\ref{algo:Basic index Query}, Algorithm~\ref{algo:advanced index query} based on offline construction and online querying have a performance advantage of 2 to 3 orders of magnitude. Additionally, by observing the time efficiency between the two types of indices, it can be seen that the query time of the basic index structure is slightly less than that of the optimized index structure. This is because the optimized index adopts a space-for-time design principle, saving a substantial amount of space at the cost of a slight compromise in time performance.

\textbf{Space efficiency evaluation.} As shown in Fig.~\ref{fig:efficiency space}, the four subplots, respectively, demonstrate the variation of space usage for two types of indexes, \textbf{IHSC} and \textbf{OIHSC}, on four datasets with the change of value $k$ and the default meta-path as shown in Table.\ref{tab:parameters}. The horizontal axis represents the varying $k$ values, while the vertical axis stands for the order of magnitude of space usage in megabytes (MB).
It can be observed that \textbf{OIHSC}, when compared to \textbf{IHSC}, achieves a performance advantage of one to two orders of magnitude in terms of space storage. Moreover, as the value of $k$ increases, the amount of space optimization decreases, suggesting that the more complex the index tree is, the more space our advanced indexing structure saves.
Compared with the same types of index, as the value of $k$ increases, the space usage of the \textbf{IHSC} index monotonically decreases. This corresponds well to direct inference: as the value of $k$ increases, the set of nodes in the \textbf{IHSC} index is diminishing. However, the variation of the \textbf{IHSC} index does not possess monotonicity. Considering its feature of compressing stored vertices, it may be plausible that there might exist some subtrees of certain $k$ that store more nodes than the subtrees corresponding to larger $k$ values.

\subsection{Effectiveness evaluation}

\begin{figure}[t]
\centering
\subfigcapskip=-4pt
\subfigure[Significance w.r.t. $k$] {\label{fig:attribute_k}
\includegraphics[width=0.426\columnwidth]{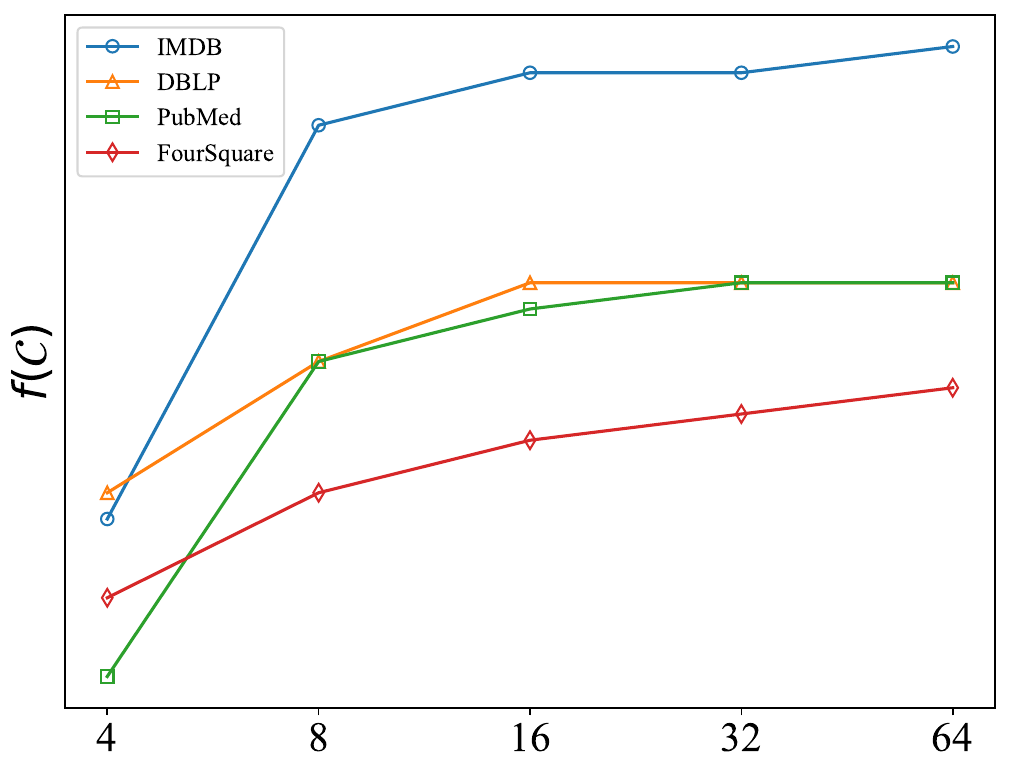}
}
\subfigure[Significance w.r.t. scale] {\label{fig:attribute_per}
\includegraphics[width=0.426\columnwidth]{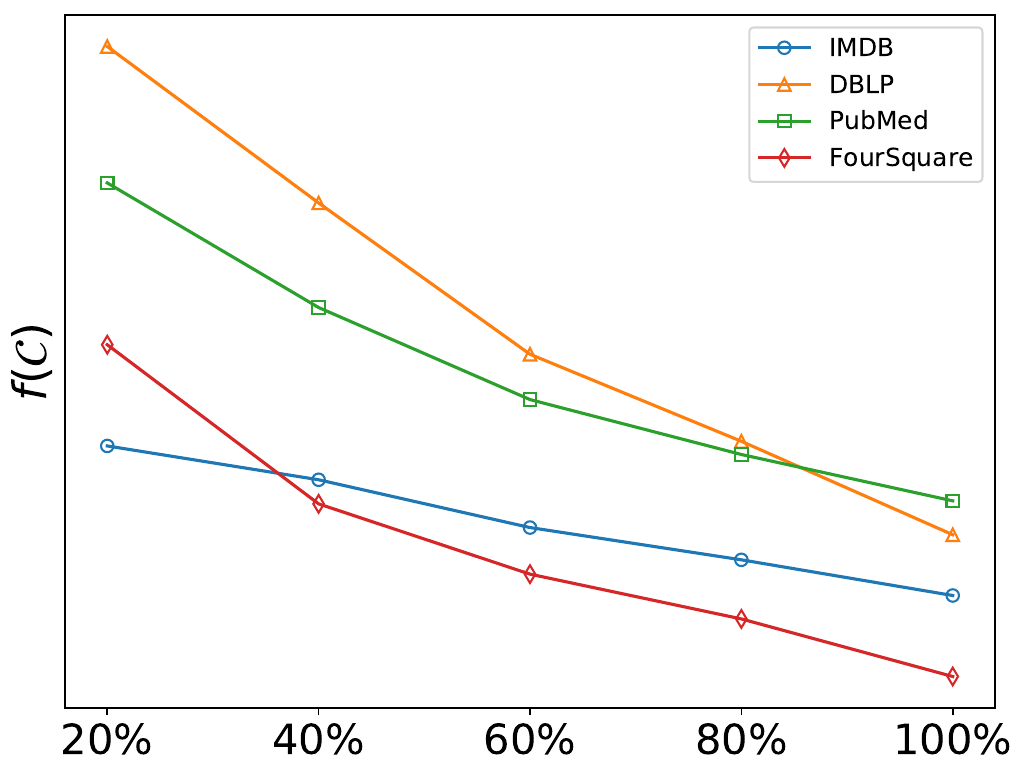}
}
\vspace{-7pt}
\caption{Significance change trend.}
\label{fig:attribute}
\vspace{-8pt}
\end{figure}

\begin{figure}[t]
\centering
\subfigcapskip=-4pt
\subfigure[QHSC] {\label{fig:scalability_QHSC}
\includegraphics[width=0.426\columnwidth]{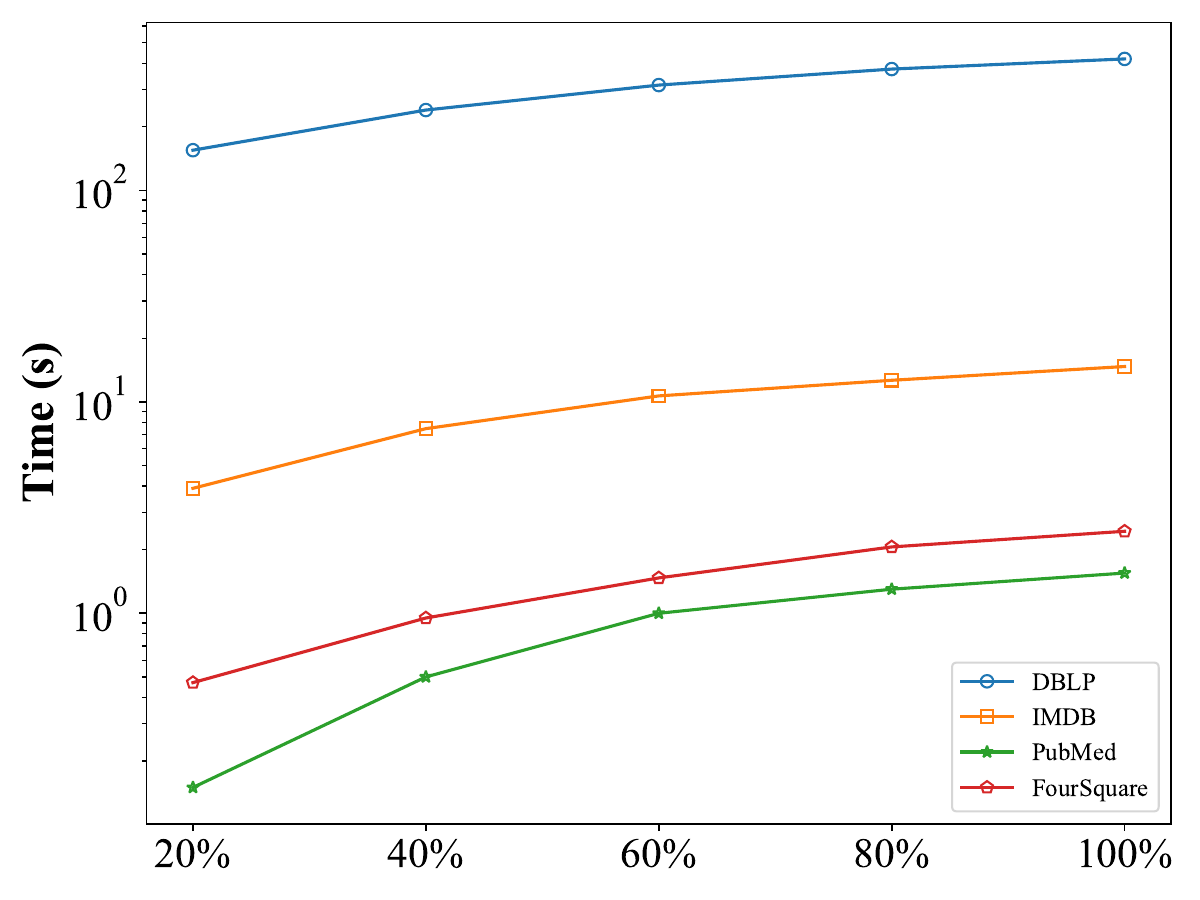}
}
\subfigure[IHSC] {\label{fig:scalability_ISAC}
\includegraphics[width=0.426\columnwidth]{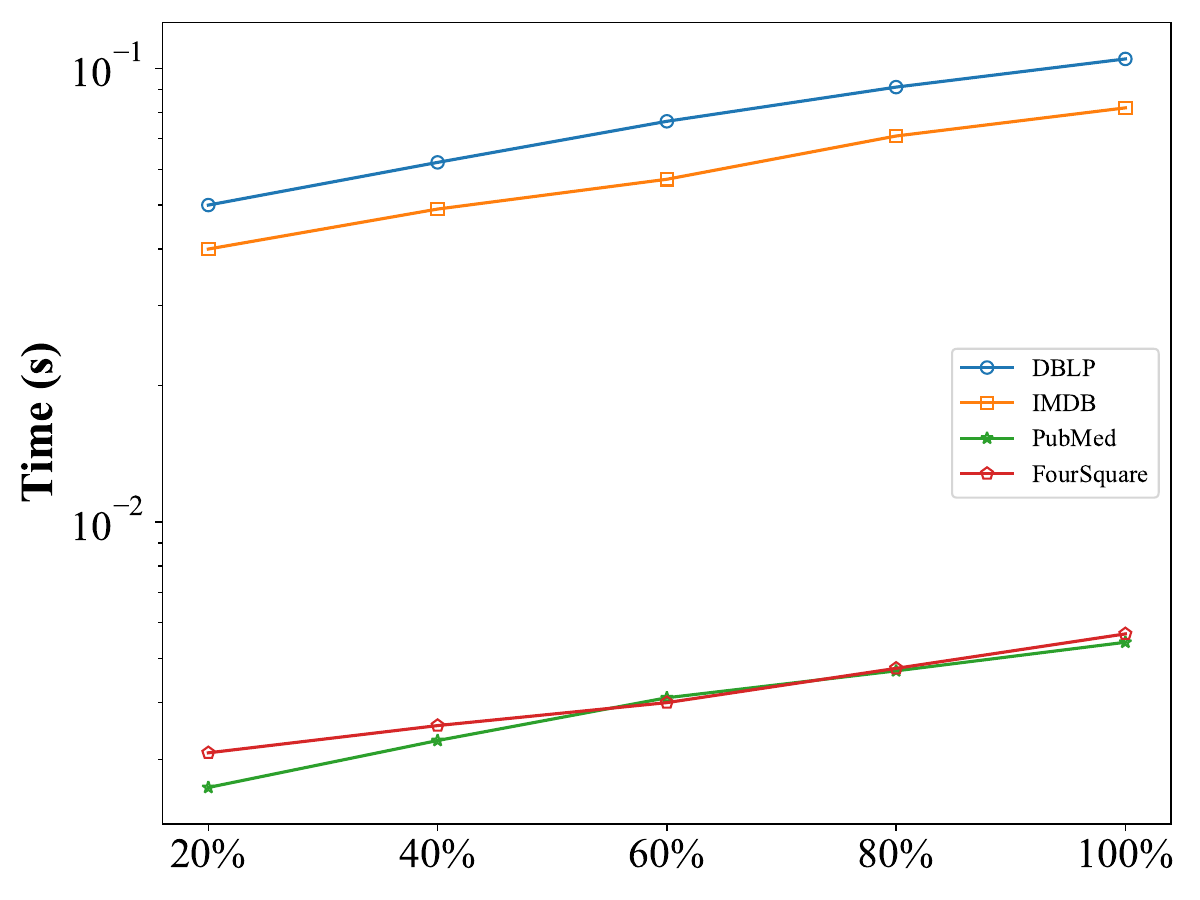}
}
\vspace{-7pt}
\caption{Scalability test for query algorithms.}
\label{fig:scalability}
\vspace{-8pt}
\end{figure}

\textbf{Scalability evaluation.} 
Fig.~\ref{fig:scalability} illustrates the scalability of the Algorithm~\ref{algo:QHSC} and Algorithm \ref{algo:Basic index Query}. We randomly selected 20\%, 40\%, 60\%, 80\%, and 100\% of the vertices as input graphs. During the construction of these subgraphs, edges connecting vertices that are both included in the subgraph are also included. From the test results presented in Fig.~\ref{fig:scalability_QHSC}, our algorithm demonstrates satisfactory performance across all scale levels for the four datasets, and as the scale increases, the efficiency of the queries also improves incrementally. The trend in time variation is quite evident.
On the other hand, the efficiency evaluation based on IHSC, as shown in Fig.~\ref{fig:scalability_ISAC}, does not experience a significant impact as the graph scale increases. This is an advantage of index-based query methods; once the index is well established, the query efficiency is extremely high.

\textbf{Trend of significance change evaluation.} Fig.~\ref{fig:attribute_k} and Fig.~\ref{fig:attribute_per} respectively demonstrate the evolution of the community significance $f(\mathcal{C})$ of the HSCs with the variations in the value of $k$ and the amount of data in the datasets. The experimental results suggest that the change in community value is non-increasing as $k$ increases. This is due to the fact that when $k$ gets larger, some vertices in the graph that do not meet the structural cohesiveness requirement are removed. For a graph with significance, the community value would either remain the same or increase when the vertices are removed. On the other hand, a positive correlation can be discerned between the community significance and the metric of structural cohesiveness. Thus, when querying an HSC, it is feasible to elevate the community value level by appropriately raising the value of $k$, thus enhancing the constraint of $\alpha(v, S_G)$ on structural cohesion. Similarly, as the size of the utilized graph data increases, the size of the discovered community either remains the same or increases. Based on \textit{Definition}~\ref{definition:Attribute of Community}, the community significance is the minimum value of all vertices within the community; therefore, as the number of vertices grows, its value stays the same or decreases.

\subsection{Case Study}
\begin{figure}
    \centering
    \subfigure[Example $\mathcal{C}_1$ with $f(\mathcal{C}_1)=4.3$]{\label{fig:case study gg}
    \includegraphics[width=0.45\columnwidth]{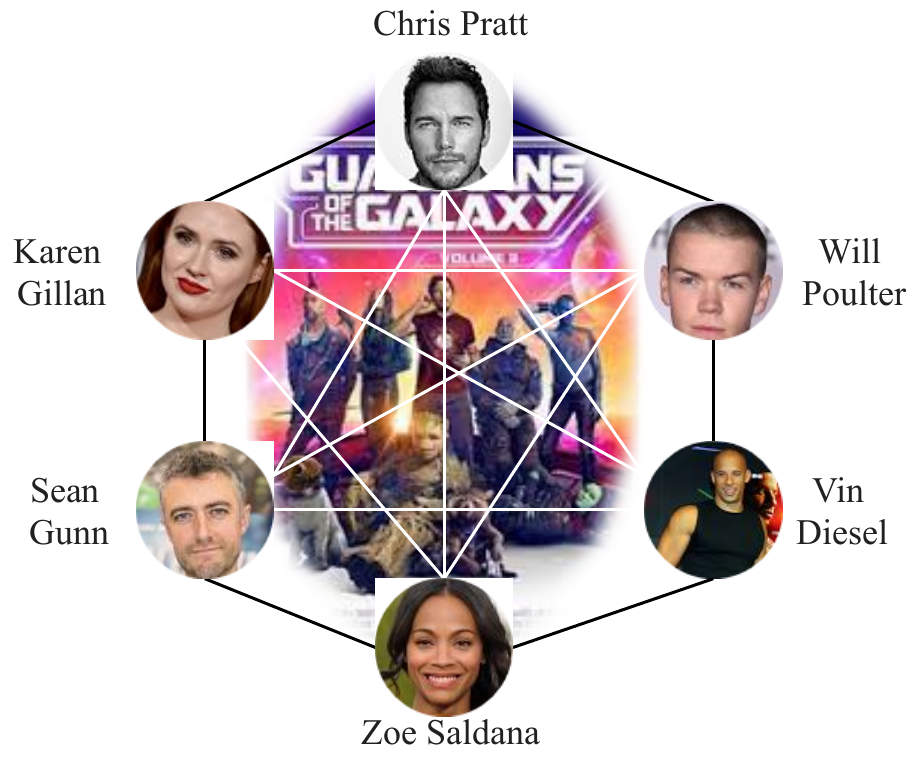}
    }
    \subfigure[$\mathcal{C}_2$ with $f(\mathcal{C}_1)=8.9$] {\label{fig:case study ta}
    \includegraphics[width=0.45\columnwidth]{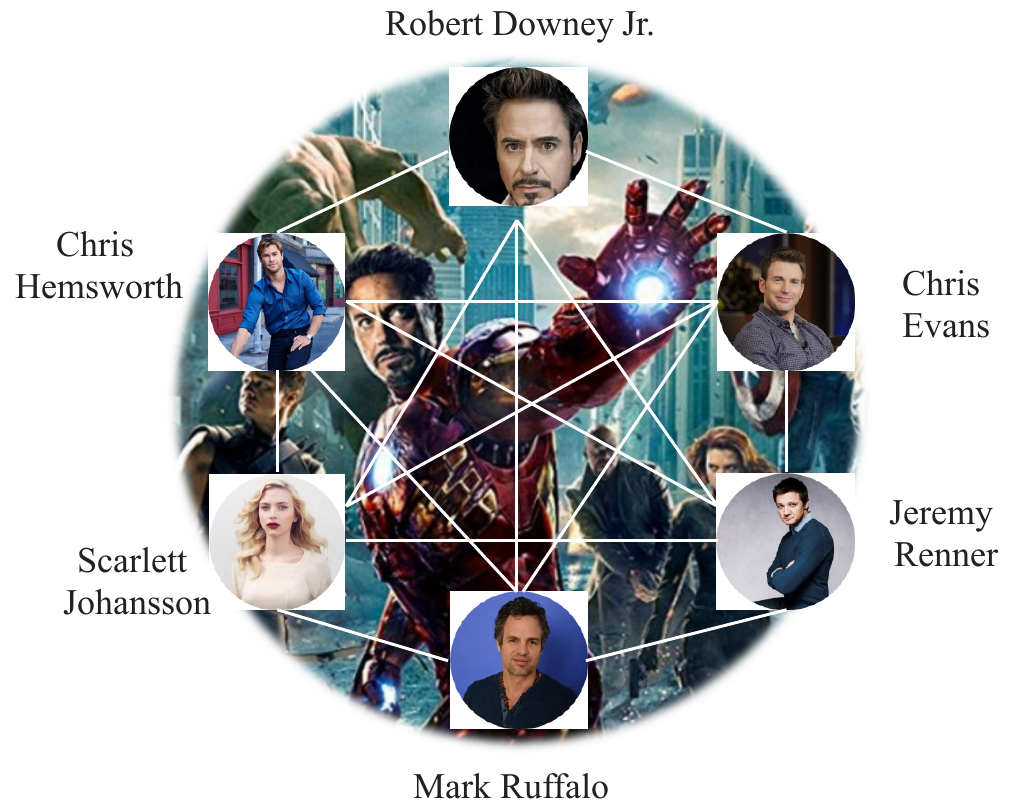}
    }
    \vspace{-3pt}
    \caption{Two HSCs in IMDB network}
    \vspace{-15pt}
    \label{fig:case study}
\end{figure}

We use a small amount of IMDB data for case analysis, letting $\mathcal{P}$={\textit{(AMA)}}, $k$=5. We select two HSCs $\mathcal{C}_1$ and $\mathcal{C}_2$ from the query nodes, which are shown in Fig.~\ref{fig:case study gg} and Fig.~\ref{fig:case study ta}. These two communities represent groups of six actors who participated in the same movie with identical structural compactness. However, the differences are $f(\mathcal{C}_1)$=4.3 and $f(\mathcal{C}_2)$=8.9, indicating that the lowest ratings among the six main actors in these two movies are 4.3 and 8.9, respectively. Clearly, the principal actors in the community $\mathcal{C}_2$ have a higher comprehensive evaluation and influence. Such significance community search is helpful for us to discover movies with a luxurious cast.

\section{Related Work}
\textbf{Community model and search.} Community search\cite{sozio2010community,cui2014local,barbieri2015efficient,huang2017community,fang2020survey} is one of the important problems in the field of graph data mining\cite{aridhi2016big}, which aims to find dense subgraphs in the network that satisfy personalized query conditions and cohesiveness \cite{yao2021efficient,liu2021efficient,dong2021butterfly}. Currently, the widely accepted classic models for measuring community cohesiveness include $k$-core\cite{seidman1983network,batagelj2003m}, $k$-truss\cite{cohen2008trusses, huang2014querying} and $k$-clique\cite{palla2005uncovering,cui2013online}. Regarding search methodologies, the primary categories are global-based search, local-based search, and index-based search. Several research groups have conducted foundational studies in these areas:
\cite{sozio2010community} introduced a global search-oriented greedy algorithm that operates in a peeling fashion, \cite{cui2014local} proposed a local expansion-based search method, \cite{barbieri2015efficient} developed an offline index structure that arranges the connected $k$-cores. 
From a structural perspective, most of the discussed studies emphasize undirected graphs, while \cite{giatsidis2013d} introduced a directed cohesiveness metric termed the $(k,l)$-core and \cite{fang2018effective} further advanced the minimum degree measure approach to cater to directed graphs.


\textbf{Attributed community search.} 
\cite{fang2016effective} explored community search on keyword attribute graphs and introduced the CL-Tree index. 
\cite{fang2017effective} addressed the community search on location attribute graphs, presenting a spatial-aware model. 
\cite{li2018persistent} investigated persistent community detection in temporal graphs. 
\cite{chen2018exploring} delved into community searches on profile attribute graphs. 
\cite{li2015influential} proposed a linear-time algorithm to identify the top j influential communities, complemented by an indexing method. Building on this foundation, \cite{chen2016efficient}\cite{bi2017optimal} advanced with reverse and local search algorithms, respectively, for enhanced efficiency. 
\cite{li2017finding} introduced an I/O-efficient approach for top-j $k$-influential communities. Moreover, to capture multi-dimensional attributes, \cite{li2018skyline} proposed a skyline community model to detect the dominant $k$-core in a d-dimensional space.
\cite{huang2017attribute} investigated the attribute-driven $k$-truss community search problem in keyword attribute graphs. 
\cite{zheng2017finding} introduced a new community model for weighted attribute graphs, termed the weighted $k$-truss community. 
\cite{guo2021multi} proposed a novel method for a multi-attributed joint community search on road social networks.

\textbf{Heterogeneous community search.} 
In recent years, some researchers have extended the community search problem from homogeneous graphs to HINs, 
\cite{hu2019discovering} explored clique based community search over HINs and \cite{dong2021butterfly} studied label graphs with butterfly-core.
\cite{zhang2021pareto} introduced a pareto-optimal community search model over bipartite graphs, 
\cite{jian2020effective} processed the community search problem over dynamic HINs.
\cite{fang2020effective, jiang2022effective, fang2022cohesive} proposed the \((k, \mathcal{P})\)-core to characterize the cohesiveness level of communities in heterogeneous graphs.
Integrating attribute and heterogeneous structural, \cite{qiao2021keyword} conducted research on HIN by incorporating keywords and recently \cite{zhou2023influential} introduced influence into the community search problem on heterogeneous graphs. These make the results of community search more accurate and closer to real-world applications. 

\section{Conclusions}
In this paper, we study the problem of significant-attributed community search in HINs (SACH), which aims to unravel closely connected vertices of the anchor-type with high importance through multiple semantic relationships. In particular, we introduce a novel community model called heterogeneous significant community (HSC). To search HSCs, we first develop online algorithms by exploiting both segmented-based meta-path expansion and significance increment. Specially, a solution space reuse strategy based on structural nesting is designed to boost the efficiency. Additionally, a two-level index is further devised to support searching HSCs in optimal time, and a space-efficient compact index is proposed. Experimental results on real large HINs demonstrate that our solutions are effective and efficient for searching HSCs.

\section{Acknowledgment} 
We would like to express our gratitude to Yixiang Fang and his research team for laying the groundwork with their outstanding research achievements in this field. Additionally, the open-source code provided by Yangqin Jiang\cite{jiang2022effective} of the team has greatly facilitated our research. Fangda Guo is the corresponding author.

\bibliographystyle{IEEEtran}
\clearpage
\bibliography{IEEEabrv,ref}

\begin{thebibliography}{10}
\providecommand{\url}[1]{#1}
\csname url@samestyle\endcsname
\providecommand{\newblock}{\relax}
\providecommand{\bibinfo}[2]{#2}
\providecommand{\BIBentrySTDinterwordspacing}{\spaceskip=0pt\relax}
\providecommand{\BIBentryALTinterwordstretchfactor}{4}
\providecommand{\BIBentryALTinterwordspacing}{\spaceskip=\fontdimen2\font plus
\BIBentryALTinterwordstretchfactor\fontdimen3\font minus
  \fontdimen4\font\relax}
\providecommand{\BIBforeignlanguage}[2]{{%
\expandafter\ifx\csname l@#1\endcsname\relax
\typeout{** WARNING: IEEEtran.bst: No hyphenation pattern has been}%
\typeout{** loaded for the language `#1'. Using the pattern for}%
\typeout{** the default language instead.}%
\else
\language=\csname l@#1\endcsname
\fi
#2}}
\providecommand{\BIBdecl}{\relax}
\BIBdecl

\bibitem{sun2012mining}
Y.~Sun and J.~Han, \emph{Mining heterogeneous information networks: principles
  and methodologies}.\hskip 1em plus 0.5em minus 0.4em\relax Morgan \& Claypool
  Publishers, 2012.

\bibitem{cui2014local}
W.~Cui, Y.~Xiao, H.~Wang, and W.~Wang, ``Local search of communities in large
  graphs,'' in \emph{SIGMOD}, 2014, pp. 991--1002.

\bibitem{huang2015approximate}
X.~Huang, L.~V. Lakshmanan, J.~X. Yu, and H.~Cheng, ``Approximate closest
  community search in networks,'' \emph{PVLDB}, vol.~9, no.~4, pp. 276--287,
  2015.

\bibitem{fang2016effective}
Y.~Fang, R.~Cheng, S.~Luo, and J.~Hu, ``Effective community search for large
  attributed graphs,'' \emph{PVLDB}, vol.~9, no.~12, pp. 1233--1244, 2016.

\bibitem{fang2017effective}
Y.~Fang, R.~Cheng, X.~Li, S.~Luo, and J.~Hu, ``Effective community search over
  large spatial graphs.'' \emph{PVLDB}, vol.~10, no.~6, pp. 709--720, 2017.

\bibitem{li2018skyline}
R.~Li, L.~Qin, F.~Ye, J.~X. Yu, X.~Xiao, N.~Xiao, and Z.~Zheng, ``Skyline
  community search in multi-valued networks,'' in \emph{SIGMOD}, 2018, pp.
  457--472.

\bibitem{han2022data}
J.~Han, J.~Pei, and H.~Tong, \emph{Data mining: concepts and techniques}.\hskip
  1em plus 0.5em minus 0.4em\relax Morgan kaufmann, 2022.

\bibitem{li2015influential}
R.~Li, L.~Qin, J.~X. Yu, and R.~Mao, ``Influential community search in large
  networks,'' \emph{PVLDB}, vol.~8, no.~5, pp. 509--520, 2015.

\bibitem{huang2017attribute}
X.~Huang and L.~V. Lakshmanan, ``Attribute-driven community search,''
  \emph{PVLDB}, vol.~10, no.~9, pp. 949--960, 2017.

\bibitem{li2017finding}
R.-H. Li, L.~Qin, J.~X. Yu, and R.~Mao, ``Finding influential communities in
  massive networks,'' \emph{The VLDB Journal}, vol.~26, pp. 751--776, 2017.

\bibitem{luo2020efficient}
J.~Luo, X.~Cao, X.~Xie, Q.~Qu, Z.~Xu, and C.~S. Jensen, ``Efficient
  attribute-constrained co-located community search,'' in \emph{ICDE}, 2020,
  pp. 1201--1212.

\bibitem{zhou2023influential}
Y.~Zhou, Y.~Fang, W.~Luo, and Y.~Ye, ``Influential community search over large
  heterogeneous information networks,'' \emph{Proceedings of the VLDB
  Endowment}, vol.~16, no.~8, pp. 2047--2060, 2023.

\bibitem{hu2019discovering}
J.~Hu, R.~Cheng, K.~C.-C. Chang, A.~Sankar, Y.~Fang, and B.~Y. Lam,
  ``Discovering maximal motif cliques in large heterogeneous information
  networks,'' in \emph{ICDE}.\hskip 1em plus 0.5em minus 0.4em\relax IEEE,
  2019, pp. 746--757.

\bibitem{fang2020effective}
Y.~Fang, Y.~Yang, W.~Zhang, X.~Lin, and X.~Cao, ``Effective and efficient
  community search over large heterogeneous information networks,''
  \emph{Proceedings of the VLDB Endowment}, vol.~13, no.~6, pp. 854--867, 2020.

\bibitem{qiao2021keyword}
L.~Qiao, Z.~Zhang, Y.~Yuan, C.~Chen, and G.~Wang, ``Keyword-centric community
  search over large heterogeneous information networks,'' in
  \emph{International Conference on Database Systems for Advanced
  Applications}.\hskip 1em plus 0.5em minus 0.4em\relax Springer, 2021, pp.
  158--173.

\bibitem{jiang2022effective}
Y.~Jiang, Y.~Fang, C.~Ma, X.~Cao, and C.~Li, ``Effective community search over
  large star-schema heterogeneous information networks,'' \emph{Proceedings of
  the VLDB Endowment}, vol.~15, no.~11, pp. 2307--2320, 2022.

\bibitem{fang2022cohesive}
Y.~Fang, K.~Wang, X.~Lin, and W.~Zhang, \emph{Cohesive Subgraph Search Over
  Large Heterogeneous Information Networks}.\hskip 1em plus 0.5em minus
  0.4em\relax Springer, 2022.

\bibitem{zhang2021pareto}
Y.~Zhang, K.~Wang, W.~Zhang, X.~Lin, and Y.~Zhang, ``par on large bipartite
  graphs,'' in \emph{Proceedings of the 30th ACM International Conference on
  Information \& Knowledge Management}, 2021, pp. 2647--2656.

\bibitem{sun2011pathsim}
Y.~Sun, J.~Han, X.~Yan, P.~S. Yu, and T.~Wu, ``Pathsim: Meta path-based top-k
  similarity search in heterogeneous information networks,'' \emph{Proceedings
  of the VLDB Endowment}, vol.~4, no.~11, pp. 992--1003, 2011.

\bibitem{shi2016survey}
C.~Shi, Y.~Li, J.~Zhang, Y.~Sun, and S.~Y. Philip, ``A survey of heterogeneous
  information network analysis,'' \emph{TKDE}, vol.~29, no.~1, pp. 17--37,
  2016.

\bibitem{maas2011learning}
A.~Maas, R.~E. Daly, P.~T. Pham, D.~Huang, A.~Y. Ng, and C.~Potts, ``Learning
  word vectors for sentiment analysis,'' in \emph{Proceedings of the 49th
  annual meeting of the association for computational linguistics: Human
  language technologies}, 2011, pp. 142--150.

\bibitem{dong2017metapath2vec}
Y.~Dong, N.~V. Chawla, and A.~Swami, ``metapath2vec: Scalable representation
  learning for heterogeneous networks,'' in \emph{Proceedings of the 23rd ACM
  SIGKDD international conference on knowledge discovery and data mining},
  2017, pp. 135--144.

\bibitem{lu2011pubmed}
Z.~Lu, ``Pubmed and beyond: a survey of web tools for searching biomedical
  literature,'' \emph{Database}, vol. 2011, p. baq036, 2011.

\bibitem{ley2009dblp}
M.~Ley, ``Dblp: some lessons learned,'' \emph{Proceedings of the VLDB
  Endowment}, vol.~2, no.~2, pp. 1493--1500, 2009.

\bibitem{cho2011friendship}
E.~Cho, S.~A. Myers, and J.~Leskovec, ``Friendship and mobility: user movement
  in location-based social networks,'' in \emph{SIGKDD}, 2011, pp. 1082--1090.

\bibitem{borzsony2001skyline}
S.~Borzsony, D.~Kossmann, and K.~Stocker, ``The skyline operator,'' in
  \emph{ICDE}, 2001, pp. 421--430.

\bibitem{sozio2010community}
M.~Sozio and A.~Gionis, ``The community-search problem and how to plan a
  successful cocktail party,'' in \emph{SIGKDD}, 2010, pp. 939--948.

\bibitem{barbieri2015efficient}
N.~Barbieri, F.~Bonchi, E.~Galimberti, and F.~Gullo, ``Efficient and effective
  community search,'' \emph{DMKD}, vol.~29, no.~5, pp. 1406--1433, 2015.

\bibitem{huang2017community}
X.~Huang, L.~V. Lakshmanan, and J.~Xu, ``Community search over big graphs:
  Models, algorithms, and opportunities,'' in \emph{ICDE}.\hskip 1em plus 0.5em
  minus 0.4em\relax IEEE, 2017, pp. 1451--1454.

\bibitem{fang2020survey}
Y.~Fang, X.~Huang, L.~Qin, Y.~Zhang, W.~Zhang, R.~Cheng, and X.~Lin, ``A survey
  of community search over big graphs,'' \emph{The VLDB Journal}, vol.~29, pp.
  353--392, 2020.

\bibitem{aridhi2016big}
S.~Aridhi and E.~M. Nguifo, ``Big graph mining: Frameworks and techniques,''
  \emph{Big Data Research}, vol.~6, pp. 1--10, 2016.

\bibitem{yao2021efficient}
K.~Yao and L.~Chang, ``Efficient size-bounded community search over large
  networks,'' \emph{Proceedings of the VLDB Endowment}, vol.~14, no.~8, pp.
  1441--1453, 2021.

\bibitem{liu2021efficient}
B.~Liu, F.~Zhang, W.~Zhang, X.~Lin, and Y.~Zhang, ``Efficient community search
  with size constraint,'' in \emph{ICDE}.\hskip 1em plus 0.5em minus
  0.4em\relax IEEE, 2021, pp. 97--108.

\bibitem{dong2021butterfly}
Z.~Dong, X.~Huang, G.~Yuan, H.~Zhu, and H.~Xiong, ``Butterfly-core community
  search over labeled graphs,'' \emph{arXiv preprint arXiv:2105.08628}, 2021.

\bibitem{seidman1983network}
S.~B. Seidman, ``Network structure and minimum degree,'' \emph{Social
  networks}, vol.~5, no.~3, pp. 269--287, 1983.

\bibitem{batagelj2003m}
V.~Batagelj and M.~Zaversnik, ``An o(m) algorithm for cores decomposition of
  networks,'' \emph{CoRR, cs.DS/0310049}, 2003.

\bibitem{cohen2008trusses}
J.~Cohen, ``Trusses: Cohesive subgraphs for social network analysis,''
  \emph{National security agency technical report}, vol.~16, no. 3.1, pp.
  1--29, 2008.

\bibitem{huang2014querying}
X.~Huang, H.~Cheng, L.~Qin, W.~Tian, and J.~X. Yu, ``Querying k-truss community
  in large and dynamic graphs,'' in \emph{SIGMOD}, 2014, pp. 1311--1322.

\bibitem{palla2005uncovering}
G.~Palla, I.~Der{\'e}nyi, I.~Farkas, and T.~Vicsek, ``Uncovering the
  overlapping community structure of complex networks in nature and society,''
  \emph{nature}, vol. 435, no. 7043, pp. 814--818, 2005.

\bibitem{cui2013online}
W.~Cui, Y.~Xiao, H.~Wang, Y.~Lu, and W.~Wang, ``Online search of overlapping
  communities,'' in \emph{SIGMOD}, 2013, pp. 277--288.

\bibitem{giatsidis2013d}
C.~Giatsidis, D.~M. Thilikos, and M.~Vazirgiannis, ``D-cores: measuring
  collaboration of directed graphs based on degeneracy,'' \emph{Knowledge and
  information systems}, vol.~35, no.~2, pp. 311--343, 2013.

\bibitem{fang2018effective}
Y.~Fang, Z.~Wang, R.~Cheng, H.~Wang, and J.~Hu, ``Effective and efficient
  community search over large directed graphs,'' \emph{IEEE Transactions on
  Knowledge and Data Engineering}, vol.~31, no.~11, pp. 2093--2107, 2018.

\bibitem{li2018persistent}
R.-H. Li, J.~Su, L.~Qin, J.~X. Yu, and Q.~Dai, ``Persistent community search in
  temporal networks,'' in \emph{ICDE}.\hskip 1em plus 0.5em minus 0.4em\relax
  IEEE, 2018, pp. 797--808.

\bibitem{chen2018exploring}
Y.~Chen, Y.~Fang, R.~Cheng, Y.~Li, X.~Chen, and J.~Zhang, ``Exploring
  communities in large profiled graphs,'' \emph{TKDE}, vol.~31, no.~8, pp.
  1624--1629, 2018.

\bibitem{chen2016efficient}
S.~Chen, R.~Wei, D.~Popova, and A.~Thomo, ``Efficient computation of importance
  based communities in web-scale networks using a single machine,'' in
  \emph{CIKM}, 2016, pp. 1553--1562.

\bibitem{bi2017optimal}
F.~Bi, L.~Chang, X.~Lin, and W.~Zhang, ``An optimal and progressive approach to
  online search of top-k influential communities,'' \emph{arXiv preprint
  arXiv:1711.05857}, 2017.

\bibitem{zheng2017finding}
Z.~Zheng, F.~Ye, R.-H. Li, G.~Ling, and T.~Jin, ``Finding weighted k-truss
  communities in large networks,'' \emph{Information Sciences}, vol. 417, pp.
  344--360, 2017.

\bibitem{guo2021multi}
F.~Guo, Y.~Yuan, G.~Wang, X.~Zhao, and H.~Sun, ``Multi-attributed community
  search in road-social networks,'' in \emph{2021 IEEE 37th International
  Conference on Data Engineering (ICDE)}.\hskip 1em plus 0.5em minus
  0.4em\relax IEEE, 2021, pp. 109--120.

\bibitem{jian2020effective}
X.~Jian, Y.~Wang, and L.~Chen, ``Effective and efficient relational community
  detection and search in large dynamic heterogeneous information networks,''
  \emph{Proceedings of the VLDB Endowment}, vol.~13, no.~10, pp. 1723--1736,
  2020.

\end{thebibliography}
\end{document}